\documentclass[11pt]{article}
\usepackage{mathtools}
\usepackage[top=2cm, left=2cm, right=2cm, bottom=2cm	]{geometry} 
\usepackage{amsmath,amsthm,amssymb}
\usepackage{amsmath}               
  {
      \theoremstyle{plain}
      
  }
\makeatletter
\newcounter{savesection}
\newcounter{apdxsection}
\renewcommand\appendix{\par
  \setcounter{savesection}{\value{section}}%
  \setcounter{section}{\value{apdxsection}}%
  \setcounter{subsection}{0}%
  \gdef\thesection{\@Alph\c@section}}
\newcommand\unappendix{\par
  \setcounter{apdxsection}{\value{section}}%
  \setcounter{section}{\value{savesection}}%
  \setcounter{subsection}{0}%
  \gdef\thesection{\@arabic\c@section}}
\makeatother

\usepackage{calrsfs}
\DeclareMathAlphabet{\pazocal}{OMS}{zplm}{m}{n}
\usepackage{bm}
\usepackage{listings}
\usepackage{natbib} 
\usepackage[ruled,vlined]{algorithm2e}
\usepackage{float}
\usepackage{mathtools}

\usepackage[nodisplayskipstretch]{setspace}
\usepackage[dvipsnames]{xcolor}
\usepackage{graphicx}
\usepackage[toc,page]{appendix}
\usepackage[nodisplayskipstretch]{setspace}
\usepackage{subcaption}

\usepackage{hyperref}
\usepackage{authblk}

\newcommand\myshade{85}
\colorlet{mylinkcolor}{violet}
\colorlet{mycitecolor}{YellowOrange}
\colorlet{myurlcolor}{Aquamarine}

\hypersetup{
  linkcolor  = mylinkcolor!\myshade!black,
  citecolor  = mycitecolor!\myshade!black,
  urlcolor   = myurlcolor!\myshade!black,
  colorlinks = true,
}

\newcommand{\dd}{\mathrm{d}}

\newcommand{\Mb}{\pazocal{M}}
\newcommand{\Hb}{\pazocal{H}}

\begin{document}

\title{Interim recruitment prediction for multi-centre clinical trials} %
\author[1]{Szymon Urbas\footnote{Email: \texttt{s.urbas@lancaster.ac.uk}, Address: STOR-i Centre for Doctoral Training, Lancaster University, Lancaster, United Kingdom}}
\author[2]{Chris Sherlock}
\author[3]{Paul Metcalfe}
\affil[1]{STOR-i Centre for Doctoral Training, Lancaster University, Lancaster, United Kingdom}
\affil[2]{Department of Mathematics and Statistics, Lancaster University, Lancaster, United Kingdom}
\affil[3]{AstraZeneca, Cambridge, United Kingdom}
\date{}

\maketitle
\begin{abstract}
We introduce a general framework for monitoring, modelling, and predicting the recruitment to multi-centre clinical trials. The work is motivated by overly optimistic and narrow prediction intervals produced by existing time-homogeneous recruitment models for multi-centre recruitment. We first present two tests for detection of decay in recruitment rates, together with a power study. We then introduce a model based on the inhomogeneous Poisson process with monotonically decaying intensity, motivated by recruitment trends observed in oncology trials. The general form of the model permits adaptation to any parametric curve-shape. A general method for constructing sensible parameter priors is provided and Bayesian model averaging is used for making predictions which account for the uncertainty in both the parameters and the model. The validity of the method and its robustness to misspecification are tested using simulated datasets. The new methodology is then applied to oncology trial data, where we make interim accrual predictions, comparing them to those obtained by existing methods, and indicate where unexpected changes in the accrual pattern occur.
\end{abstract}

\section{Introduction}
\label{sec:intro}
Efficiently recruiting patients to clinical trials is a critical factor in running clinical trials and hence delivering new medicines to patients as quickly as possible. Late-stage clinical trials are commonly run across many sites, and successfully managing and running trials and subsequent processes requires accurate forecasts of trial recruitment. 

Early recruitment rates can be high, for example, because patients with the required condition are already available, and rates can then drop once these patients have been recruited. Deterministic approaches and ad hoc techniques may yield simplified and, often, overly optimistic recruitment timelines, a phenomenon thus dubbed \emph{Lasagna's Law} \citep{Lasa1979}. For example, 48\% of centres studied by \cite{Tuft2013}  failed to enrol the required number of patients in the time originally allocated, leading to extensions of the recruitment timelines and the need to bring more centres into the study, which itself is a costly process. The timelines are usually pushed to nearly twice the originally proposed plan. The most frequent reason for trial discontinuation appears to be poor recruitment; out of 253 discontinued trials studied in \cite{KaEl2014}, 101 were terminated due to under-recruitment.

This motivates the need for robust statistical methods for modelling and predicting the recruitment to clinical trials at site-level. Early detections of possible centre underperformance may allow practitioners to swiftly intervene in the operations. It can also provide realistic timelines for the completion of different stages of the trials. 

In this work, we  introduce a novel flexible framework for effectively modelling and predicting patient recruitment. We will focus on the oncology therapeutic area as it is known for sparse enrolments whose patterns are not sufficiently captured by the state-of-the-art methods \cite[]{AnFe2007,LaTa2018}. Our framework utilises time-varying recruitment rates whilst also permitting variation between recruitment centres. Inference is based on the set of known centre initiation times to date, whilst the prediction is conditional on a set of future initiation times. Past initiation times are known, but typically, whilst there is a plan for future initiation times along with potential contingencies, the actual times are not known precisely in advance. The proposed methodology can be used  with user-specified initiation schedules to facilitate the choice between different initiation-time scenarios, or it can be combined with a centre-initiation model. Predictions of future recruitment incorporate parameter and model uncertainty, which is essential when data are limited. 

Existing methods for predicting recruitment to clinical trials are overviewed in Section \ref{sec:meth}. Section \ref{sec:detect} outlines methods for detecting recruitment rate decay in the multi-centre recruitment setting along with result of a Monte Carlo power study. Section \ref{sec:model} introduces the flexible modelling framework and Section \ref{sec:Bayes} presents a general method for choosing sensible Bayesian parameter priors, along with an appropriate posterior sampling method and diagnostics. A simulation study is presented in Section \ref{sec:sim}, illustrating the fitting of the model, model validation and forecasting recruitment using Bayesian model-averaging. In Section \ref{sec:data} the model is fitted to an oncology dataset, and this is followed by a discussion in Section \ref{sec:disc}.

\section{Existing methods}\label{sec:meth}

The first statistical modelling framework for clinical trial recruitment was introduced in \cite{Lee1983}, where the recruitment was assumed to be a constant-rate Poisson process, leading to tractable inference based on interim data. \cite{WiBi1987} built on the model by considering Bayesian inference with conjugate priors. \cite{GaSi2008} and \cite{JiSi2015} further explored the effects various prior densities can have on predictions. Time-inhomogeneous accrual was first considered in \cite{PiPa1987}, where the aggregated accrual across all sites was modelled as an inhomogeneous Poisson process with intensity $\lambda(t)=\zeta(1-\exp\{-\kappa t\})$, $\zeta,\kappa>0$. \cite{ZhLo2010} took a non-parametric approach, using B-splines to model the trends in accrual and using the intensity value at the census time for predictions. \cite{TaKo2012} proposed a Poisson model with a piece-wise linear intensity which captured aspects of recruitment such as slow initial recruitment and a spike in recruitment close to the end of the trial. For a more thorough review of these as well as other methods see \cite{HeGe2015}. Accrual-only modelling methods do not consider the effect that initiating new centres can have on recruitment trends. For that reason, we shall focus on methods which can take advantage of centre-specific recruitment data.

\cite{AnFe2007} introduced the Poisson-gamma (PG) model of recruitment in a multi-centre setting, with the main appeal being the use of random effects for the  recruitment rates of centres, providing a tractable, data-driven prior predictive distribution for recruitment in yet-unopened centres. The model consists of $C$ centres, each recruiting $N_c$ patients over $\tau_c$ days, $c=1,\ldots,C$. The framework makes the following distributional assumptions,
\begin{align}\begin{aligned}
\lambda_c&\sim\mbox{Gamma}\left(\alpha,\alpha/\phi\right),\\
N_c|\lambda_c&\sim\mbox{Pois}\left(\lambda_c\tau_c\right),\end{aligned}\qquad c=1,\ldots,C.\label{eqn:anis}
\end{align}
The random effect $\lambda_c$ is the \emph{recruitment rate} for centre $c$. The rates, and thus the centre recruitments, are assumed to be independent conditional on $\alpha$ and $\phi$. There are, however, several caveats with the approach taken. The paper advocates using the Empirical Bayes approach, that is, maximum likelihood estimation for the hierarchical parameters $(\alpha,\phi)$ followed by re-estimation of the distribution of random effect $\lambda_c$ given $\alpha$, $\phi$ and $n_c$, for each centre. A method for obtaining the uncertainty in the hierarchical $(\alpha,\phi)$ parameters is provided, but this uncertainty is not accounted for when making predictions, leading to overly confident prediction intervals. However, the main issue which could result from employing the model arises from the strong assumption of time-homogeneity of centre recruitments, which can lead to underestimations of the time to completion. 

\begin{figure}
\centering
\includegraphics[width = 0.6\textwidth]{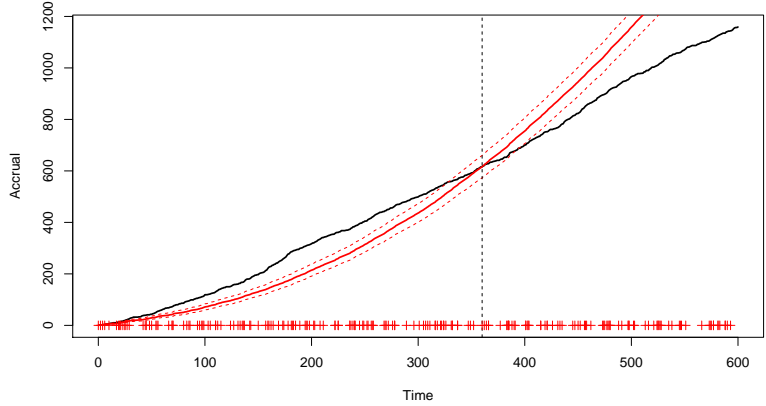}
\caption{Accrual (black, solid) with the predictive mean (red, solid) and 95\% prediction bands (red,dashed), based on the PG model (\ref{eqn:anis}) with the census time marked by the vertical, dashed line.}\label{fig:anis}
\end{figure}

Figure \ref{fig:anis} shows the accrual in a simulated trial where the rates gradually decay with time as well as the predictive distribution of the PG model fitted at a census time of three-fifths of the total length of the study; the initiation day for each centre is marked. The accrual appears to follow a straight line which could initially suggest using a time-homogeneous model. However, new centres are constantly being iniatated so that a constant recruitment rate for each centre leads to an upward arching trend in accrual. This is encapsulated by the fitted predictive. Here the accrual is initially badly underestimated and then grossly overestimated after the census time. The apparent ``matching'' at the census time is due to predictions using re-estimated random-effect distributions.

\cite{LaTa2018} describes the first multi-centre recruitment model in which the rates decrease over time. The model assumes inhomogeneous Poisson for arrivals centre $c$ with an intensity of the form
\begin{align*}
\lambda_c(t) =\begin{cases}\lambda_c^o, \quad &t<t_o\\
\lambda_c^o \exp\{-\theta (t-t_o)\}, &t\geq t_o\end{cases}, 
\end{align*}
where $\lambda_c^o$ is a gamma random effect, as in (\ref{eqn:anis}),  and $t_o$ a user-specified parameter and is not estimated as part of the inference. By enforcing the specific intensity-form, the possibilities of time-homogeneous recruitments or even intensity decays with heavier tails are excluded. A more systematic alternative is to start by testing the time-homogeneity assumption.

\section{Detecting time-inhomogeneity}\label{sec:detect}
Given series of daily centre recruitment counts over the recruitment period of $\tau_c$ days, $\left\{N_c(t)\right\}_{t=1}^{\tau_c}$, $c=1,\ldots,C$, we can test the hypothesis of time-homogeneity. To detect a decay in the rate, we only need to use the sums $X^{(c)}_1 = \sum_{t=1}^{\tau_c/2} N_c(t)$ and $X^{(c)}_2 = \sum_{t=\tau_c/2 +1}^{\tau_c} N_c(t)$ ($c=1,\ldots,C$), whose expectations we denote by  $\mu^{(c)}_1$ and $\mu^{(c)}_2$ respectively.
Detecting time-inhomogeneity in a single centre can be difficult as the infrequent counts will lead to low powers of tests \citep{KrTh2004} (see also Tables \ref{tab:lrt} and \ref{tab:bst}). Thus we combine the recruitments across all centres leading to two counts: $X_1 = \sum_{c=1}^{C}X_1^{(c)}$ and $X_2= \sum_{c=1}^{C}X_2^{(c)}$, and we choose our hypotheses to be $\mathrm{H}_0:\sum_{c=1}^C\mu^{(c)}_1=\sum_{c=1}^C\mu^{(c)}_2\quad \mbox{vs}\quad \mathrm{H}_1:\sum_{c=1}^C\mu^{(c)}_1>\sum_{c=1}^C\mu^{(c)}_2$.

The tests are one-sided as we are only interested in recruitment which decays over time.  We consider tests with respect to the following assumptions:

\textbf{Assumption 1:} \emph{For each centre $c=1,\ldots,C$, the counts in the first and second halves of that centre's recruitment period are independent and have the same distribution, $X^{(c)}_1 \stackrel{d}{=} X^{(c)}_2$, with expectation $\mu^{(c)}_1$. Furthermore, the recruitments at each centre are independent of each other.}

\textbf{Assumption 2:} \emph{The patients arrive according to a Poisson process such that $X^{(c)}_1,X^{(c)}_2\sim\mbox{Pois}\left(\mu^{(c)}_1\right)$, for some $\mu^{(c)}_1$, $c=1,\ldots,C$.}

\begin{figure}
	\centering
	\includegraphics[width = 144pt]{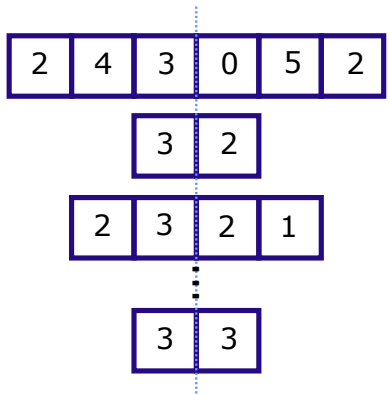}
	\caption{Count series are all centred and the sum of all the first halves is compared to the sum of second halves.}\label{fig:split}
\end{figure}

Assumption 1 implies that $X_1$ and $X_2$  must have the same distributions, with respective expectations $\mu_1 = \sum_{c=1}^{C}\mu_1^{(c)}$ and $\mu_2 = \sum_{c=1}^{C}\mu_2^{(c)}$ being equal. Assumption 2 further implies that the distributions must be Poisson. Figure \ref{fig:split} shows the construction of the quantities $X_1$ and $X_2$ by aligning the centres of the recruiting periods. The splitting of the series halfway is arbitrary, though splitting it in half (or at least close to this) would theoretically yield the highest power. It assumes that the $\tau_c$ are even. However, centres recruiting over odd numbers of days can still be used by removing the middle day observation. This reduces the power of the tests, though the reduction is negligible.

\cite{GuNg2008} offer a detailed Monte Carlo study of the different methods used for testing for a difference in means of two Poisson variables. Here, we focus on the ones most applicable to the clinical-trial recruitment setting, bearing in mind statistical power and robustness. We identified two methods: the non-parametric bootstrapped test (BST), which is powerful yet robust, and the Poisson likelihood-ratio test (LRT), which makes stronger distribution assumptions to achieve an even higher power. The BST only assumes that the counts in each day are independent and identically distributed (Assumption 1). With this assumption, resampling within each centre with replacement, from the original data would still produce a valid sample from the assumed distribution under $\mathrm{H}_0$. A large number of bootstrap samples is used to simulate the distribution of the difference in two means, which is then used to test the hypothesis. Appendix A of the Supplementary Material details the sampling procedure for obtaining the distribution and the $p$-value.

For the LRT, we require Assumption 2, which is already an underlying assumption for the model in \cite{AnFe2007}. Upon aggregation, the two sums follow Poisson distributions, that is, $X_1\sim\mbox{Pois}(\mu_1)$ and $X_2\sim\mbox{Pois}(\mu_2)$. The likelihood under the null model ($\mu_1=\mu_2$) is compared to the likelihood under the alternative two-mean model ($\mu_1>\mu_2$). Here, the likelihood function is $$L(\mu_1, \mu_2|x_1,x_2)=\frac{\mu_1^{x_1}\exp\{-\mu_1\}}{x_1!}\frac{\mu_2^{x_2}\exp\{-\mu_2\}}{x_2!},\quad \mu_1,\mu_2>0.$$
We let
\begin{align*}
T_L(x_1,x_2) = \begin{cases} 2[\log L(\hat{\mu}_1, \hat{\mu}_2|x_1,x_2)-\log L(\hat{\mu}, \hat{\mu}|x_1,x_2)],&\quad \hat{\mu}_1>\hat{\mu}_2\\
0,&\quad  \hat{\mu}_1\leq\hat{\mu}_2
\end{cases},
\end{align*}
where $\hat{\mu}$ is the MLE under the null, and $\hat{\mu_1}$ and $\hat{\mu_1}$ are the MLEs under the alternative hypothesis. Under the null, we would expect the test statistic $T_L(X_1,X_2)$ to asymptotically be zero half the time with the other half following a $\chi^2_1$ distribution \citep{RoWr1988}, When using the LRT, the simulated significance levels can differ from the pre-specified level when $\mu$ values are low. This is due to using the asymptotic $\chi^2$ distribution when calculating the $p$-value \citep{GuNg2008}.

The performance of the two tests was assessed by carrying out a Monte Carlo study. Test powers were estimated using Poisson data with different expectations and ratios, $R=\mu_2/\mu_1$. For the LRT power estimates, $5\times 10^6$ samples were used as the test itself is very computationally cheap. For the BST, $5\times 10^4$ samples were used, with each test using a bootstrapped distribution of size $10^3$. Tables \ref{tab:lrt} and \ref{tab:bst} show the results of the study. The biggest difference in powers occurs for lower expectations, with the LRT outperforming BST. It must be noted, however, that the BST only requires the data to be i.i.d. within each centre and thus is robust to violations of the Poisson assumption; if the counts within each centre are overdispersed, for example, it does not affect the Type I error.

To exemplify the usefulness of this test, we can consider an interim likelihood ratio test where the expected number of enrolments is 170. This corresponds to $E[X_1]=100$ and $R=0.7$, for example, and results in a statistical power of approximately 0.75. Considering many trials require an upward of 500 enrolments, informed decisions can be made relatively early on in the trial.

\begin{table}
	\caption{Power for likelihood-ratio test}\label{tab:lrt}
	\centering
    \begin{tabular}{r | c c c c c c}
	\hline \hline \\[0.03ex]
    $\mathbb{E}[X_1]$ &$R=1$& $R =0.9$ &  $R =0.8$ &  $R =0.7$ &  $R =0.6$ & $R =0.5$\\ [0.9ex]
    \hline
	  	 5 & 0.06 & 0.08 & 0.11 & 0.15 & 0.20 & 0.27 \\ 
  10 & 0.05 & 0.08 & 0.12 & 0.18 & 0.26 & 0.37 \\ 
  20 & 0.05 & 0.09 & 0.17 & 0.27 & 0.41 & 0.58 \\ 
  50 & 0.05 & 0.13 & 0.28 & 0.50 & 0.73 & 0.90 \\ 
  100 & 0.05 & 0.18 & 0.44 & 0.75 & 0.94 & 0.99 \\ 
  200 & 0.05 & 0.27 & 0.68 & 0.95 & 1.00 & 1.00 \\ 
	  
\hline
\end{tabular}
\end{table}

\begin{table}
	\caption{Power for non-parametric bootstrap test}\label{tab:bst}
	\centering
    \begin{tabular}{r | c c c c c c}
	\hline \hline \\[0.03ex]
    $\mathbb{E}[X_1]$ &$R=1$& $R =0.9$ &  $R =0.8$ &  $R =0.7$ &  $R =0.6$ & $R =0.5$\\ [0.9ex]
    \hline
    	  	 5 & 0.04 & 0.06 & 0.08 & 0.11 & 0.14 & 0.18 \\ 
  10 & 0.05 & 0.08 & 0.12 & 0.16 & 0.24 & 0.33 \\ 
  20 & 0.05 & 0.10 & 0.16 & 0.25 & 0.39 & 0.57 \\ 
  50 & 0.05 & 0.14 & 0.28 & 0.48 & 0.70 & 0.88 \\ 
  100 & 0.05 & 0.18 & 0.42 & 0.74 & 0.93 & 0.99 \\ 
  200 & 0.05 & 0.28 & 0.67 & 0.94 & 1.00 & 1.00 \\ 
	  
\hline
\end{tabular}
\end{table}

\section{Proposed model}\label{sec:model}

We consider a scenario of $C$ centres recruiting patients, with each centre $c$ being initiated for $\tau_c$ days. The number recruited by centre $c$ on day $t$ shall be denoted by $N_c^{(t)}$. We propose the following modelling framework for the multi-centre clinical-trial recruitment, based on the inhomogeneous Poisson process,
\begin{align*}
\lambda^o_c&\sim\mbox{Gamma}\left(\alpha,\frac{\alpha}{\phi}\right),&\quad c=1,\ldots,C,\\
N_c^{(t)}&\sim\mbox{Pois}\left(\lambda^o_c\int_{t-1}^{t}g(s;\theta)\;\mathrm{d} s\right), &\quad t=1,\ldots,\tau_c,
\end{align*}
where $g$ is a non-negative function which dictates the curve-shape of the intensity and $\theta$ is a parameter (or parameter vector) associated with the functional form. We use the $(\alpha, \phi)$ parametrisation for the hierarchical gamma distribution as it leads to orthogonality of $\alpha$ and $\phi$ in the Poisson-gamma model \citep{Huzu1950}. \emph{A priori}, $E[\lambda_c] = \phi$ and $V[\lambda_c] = \phi^2/\alpha$. For notational simplicity, we define $G(t;\theta)=\int_0^t g(s;\theta)\;\mathrm{d} s$. The likelihood contribution from centre $c$ is
\begin{align*}
\Pr(\mathbf{N}_c=\mathbf{n}_c|\lambda^o_c,\theta,\tau_c)&=\prod_{t=1}^{\tau_c} \Pr(N_c^{(t)}=n_c^{(t)}|\lambda^o_c,\theta)\\
				&= \exp\{-\lambda_c^o G(\tau_c;\theta)\}(\lambda_c^o)^{n^{(\cdot)}_{c}}\prod_{t=1}^{\tau_c} \frac{\left[G(t;\theta)-G(t-1;\theta)\right]^{n_c^{(t)}}}{n_c^{(t)}!},
\end{align*}
where $n^{(\cdot)}_{c}=\sum_{t=1}^{\tau_c} n_c^{(t)}$. Marginalising over the random-effect component gives 
\begin{align*}
\Pr(\mathbf{N}_c=\mathbf{n}_c|\alpha,\phi,\theta,\tau_c)
				=\frac{(\alpha/\phi)^\alpha\Gamma\left(\alpha+n^{(\cdot)}_{c}\right)}{\Gamma(\alpha)[G(\tau_c;\theta)+\alpha/\phi]^{\left(\alpha+n^{(\cdot)}_{c}\right)}}\prod_{t=1}^{\tau_c} \frac{\left[G(t;\theta)-G(t-1;\theta)\right]^{n_c^{(t)}}}{n_c^{(t)}!},
\end{align*}
whence the full likelihood of the model given the recruitment data is:
\begin{align}
L(\alpha,\phi,\theta|\mathbf{n},\boldsymbol\tau) & = \prod_{c=1}^{C}\Pr(\mathbf{N}_c=\mathbf{n}_c|\alpha,\phi,\boldsymbol\tau)\nonumber\\
&=\frac{(\alpha/\phi)^{C\alpha}}{\Gamma(\alpha)^C}\prod_{c=1}^C \frac{\Gamma\left(\alpha+n^{(\cdot)}_{c}\right)}{[G(\tau_c;\theta)+\alpha/\phi]^{\left(\alpha+n^{(\cdot)}_{c}\right)}}\prod_{t=1}^{\tau_c}\frac{\left[G(t;\theta)-G(t-1;\theta)\right]^{n_c^{(t)}}}{n_c^{(t)}!}.\label{eqn:lik}
\end{align}
If all the centres had been recruiting for the same amount of time, that is, $\tau_c\equiv\tau$ $\forall c$, then by fixing the integral of $g(t;\theta)$ over $\tau$ days we could introduce orthogonality between $(\alpha,\phi)$ and $\theta$ by imposing the normalisation: $
\int_0^{\tau}g(t;\theta)\;\mathrm{d} t=\tau.$
This generalises the homogeneous model with $g(t;\theta)=1$ and leads to the following factorisable likelihood,
\begin{align}
L(\alpha,\phi,\theta|\mathbf{n},\boldsymbol\tau) &=\frac{(\alpha/\phi)^{C\alpha}}{\Gamma(\alpha)^C(\tau+\alpha/\phi)^{(C\alpha+n_{\Sigma})}}\prod_{c=1}^C \Gamma\left(\alpha+n^{(\cdot)}_{c}\right)\prod_{t=1}^{\tau_c}\frac{\left[G(t;\theta)-G(t-1;\theta)\right]^{n_c^{(t)}}}{n_c^{(t)}!}\nonumber\\
&=L(\alpha,\phi|\mathbf{n},\tau)L(\theta|\mathbf{n},\boldsymbol\tau),\label{eqn:orth}
\end{align}
where $n_\Sigma = \sum_{c=1}^{C}n_c^{(\cdot)}$.

The factorisation  means that now the $\theta$ parameter describes the shape of the intensity only, and $\alpha$ and $\phi$ describe the distribution of the magnitude  of the integrated intensity, leading to a more interpretable model.

Even when centres are not all recruiting for the same length of time, we choose to impose a similar normalisation using some representative $\tau$, here $\frac{1}{C}\sum_{c=1}^{C}\tau_c$. As demonstrated empirically in Section \ref{sec:sim}, the condition leads to approximate orthogonality even when the centres are initiated uniformly throughout the study.
\subsection{Intensity curve-shape}\label{sec:int}
In this work, we will restrict our choice of curve-shape $g$ to parametric forms. The functional form of $g$ is arbitrary and the best choices may depend on the context of the problem. When working with oncology datasets, for each centre we observe low-frequency counts which seem to become even less frequent over time but with varying tail behaviours. For this reason, we chose the following curve-shape
\begin{equation}
g_\kappa(t;\theta)\propto \left(1+\frac{\theta t}{\kappa}\right)^{-\kappa},\quad t\geq 0,\;\;\theta,\kappa>0.\label{eqn:g}
\end{equation}
The proportionality is used as multiplying $g_\kappa$ by some positive constant and dividing $\phi$ by the same constant leads to the same model. The limit as $\kappa\rightarrow0$ recovers the standard PG model (\ref{eqn:anis}); and letting $\kappa\rightarrow\infty$, we obtain an exponential tail. The full (normalised) forms are then
\begin{align}
g_0(t) &\equiv 1,\label{eqn:g_0}\\
g_1(t;\theta)& = \frac{\theta(1+\theta t)^{-1}}{\log(1+\theta \tau)}\tau\label{eqn:g_1},\\
g_\kappa(t;\theta) &=\frac{\theta(1-\kappa)(1+\theta t/\kappa)^{-\kappa}}{\kappa(1+\theta \tau/\kappa)^{1-\kappa}-\kappa}\tau,\quad \kappa\notin\{0,1,\infty\},\label{eqn:g_k}\\
g_\infty(t;\theta) &= \frac{\theta \exp\{-\theta t\}}{1-\exp\{-\theta\tau\}}\tau.\label{eqn:g_inf}
\end{align}
The associated integrated forms, $G_\kappa(t;\theta)$ are provided in Appendix B of the Supplementary Material.

The flexibility of the model, however, can result in potential identifiability issues. Inference methods, such as maximum likelihood, can run into numerical instabilities when $\kappa>>1>\theta$ or $\kappa<1<<\theta$ (see Appendix B of the  Supplementary Material for details). For this reason, we recommend  restricting the choice of $\kappa$ to a discrete set of values; in this work, we use  $\{0,0.5,1,2,\infty\}$. This will be elaborated on in Section \ref{sec:bma}.

\section{Inference, diagnostics and predictions}
\label{sec:Bayes}
We aim to construct a framework which can provide reliable predictions whilst capturing  uncertainty in the estimated parameters and in the underlying model itself. We employ the Bayesian paradigm since it naturally incorporates the distribution of the random effects, $\lambda_c$, with the uncertainty in the model and the parameter values. However, we note that in some scenarios frequentist methods may be preferred and give a brief outline of how one may employ them in Appendix C of the Supplementary Material. 

Given a parametric statistical model, the Bayesian paradigm starts from a prior distribution for the parameters, here denoted $\pi_0(\alpha,\phi,\theta)$ and updates this according to some data, $y$, to provide a posterior distribution, here denoted by $\pi(\alpha,\phi,\theta|y)$. When multiple parametric models, $M_k$, $k=1,\dots,K$, are being considered, the posterior probability for model $k$, here denoted by $\pi_p(M_k|y)$, may also be calculated. Section E of the supplementary material provides more details on these quantities; see also \cite{RoCa2013} or \cite{GeCa2013}, for example. 

For the models under consideration for trial-recruitment  data, neither the posterior model probabilities nor the posteriors for the parameters for any particular model are tractable, and so we employ importance sampling to obtain Monte Carlo samples $(\alpha_m,\phi_m,\theta_m)$, $m=1,\dots,M$ from the posterior distribution for any given model, as well as an estimate of $\pi(M_k)$, $k=1,\dots,K$. Appendix D of the Supplementary Material provides further details of this method, as well as of  effective sample size (ESS), a diagnostic which indicates the reliability of the Monte Carlo estimates; see also \cite{RoCa2013} or \cite{Smit2013}.

In Sections \ref{sec:sim} and \ref{sec:data}, we carry out inference on $\tilde{\alpha} = \log\alpha$, $\tilde{\phi} = \log\phi$ and $\tilde{\theta} = \log\theta$ since analyses of trial data showed the likelihood in the log-parameters to be more symmetric about the mode, which can make sampling more efficient.
For the importance sampling proposal distribution, we use a multivariate $t$-distribution on 4 degrees of freedom, with the same mode as the posterior and the shape matrix equal to the inverse Hessian at the posterior mode.

\subsection{Prior choices}\label{sec:prior}

We base our prior specification on a maximum likelihood meta-analysis of 20 oncology clinical trial recruitment datasets. The trials studied were for seven different types of cancers: ovarian, prostate, breast, small and non-small lung, bladder and pancreatic. The number of centres ranged from 58 to 244 with a median of 140 and total enrolments ranged from 245 to 4391 with a median of 1035. In all cases, the parameter estimators were close to orthogonal justifying the use of independent priors: 
$\pi_0(\tilde{\alpha},\tilde{\phi},\tilde{\theta})=\pi_0(\tilde{\alpha})\pi_0(\tilde{\phi})\pi_0(\tilde{\theta})$.

We found that the $\alpha$ parameter does not change much from one study to another. The weakly informative prior $\tilde{\alpha}\sim N(0.2,2^2)$ sufficiently reflects the distribution of the estimated values. 

The $\phi$ parameter estimates varied by orders of magnitude between studies. The parameter reflects the mean centre recruitment and is well identified by the data; it depends upon the catchment region, type of indication and protocol, for example. For this reason, we advocate using a vague prior unless reliable expert knowledge is available. In our analyses, we used the uninformative, proper prior $\tilde{\phi}\sim U(-8,8)$.

The difference between the homogeneous (\ref{eqn:g_0}) and the inhomogeneous (\ref{eqn:g_1}, \ref{eqn:g_k}, \ref{eqn:g_inf}) models is the curve-shape parameter $\theta$. Lindley's paradox \citep{Lind1957} warns that assigning $\theta$ a vague prior can lower the posterior probabilities of the  models that use $\theta$, compared to the model with $\kappa=0$ which does not use $\theta$. To avoid the paradox we set an informative but sensible prior by considering the drop off in intensity after some time, $t_0$. We let $R_\kappa = g_\kappa(t_0;\theta)/g_\kappa(0;\theta)$ and set $R_\kappa\sim \mbox{Beta}(a,b)$ $\emph{a priori}$, with $a=b=1.1$ to indicate a lack of information, excepting that this is not a constant intensity model, since  this is covered by $\kappa=0$, and that we do not expect a 100\% drop off after a time of $t_0$ (expert opinion); here we take $t_0=4$  months. As $R_\kappa$ is a monotonic function of $\theta$, we can use a density transform to derive the corresponding prior for $\theta$. If prior information is abundant, be it in the form of historical data or expert knowledge, the beta distribution parameters can be adjusted to reflect this. Given \eqref{eqn:g}, the resulting prior density for $\tilde{\theta}$  is given in Appendix E of the Supplementary Material.

\subsection{Predictive distribution}
There are two complementary properties for which predictions might be required: the distribution of future recruitments within a set time interval, and the distribution of time until the target number of recruitments is reached. In this section, we focus on the former; details of the latter appear in  Appendix \ref{app:samp_T}.

Suppose we are interested in sampling the recruitment, denoted $N^+_c$, at some day $t^+$ by centre $c$. Given samples from the parameter posteriors, we can sample exactly from the posterior predictive for $N^+_c$ by exploiting the Poisson-gamma conjugacy of the random-effect distribution. The posterior distribution for the $\lambda^o_c$ random effect for centre $c$ is
\begin{equation}
\lambda^o_c|\alpha, \phi, \theta,\mathbf{n}_c,\tau_c \sim \mbox{Gamma}\left(\alpha+n^{(\cdot)}_{c}, \alpha/\phi+G(\tau_c;\theta)\right)=\mbox{Gamma}\left(\alpha^*_{c}, \frac{\alpha_c^*}{\phi^*_c}\right),\label{eqn:re-est}
\end{equation}
where $\alpha^*_c = \alpha+n^{(\cdot)}$ and $\phi^*_c = \phi\times\left(\frac{\alpha+n_c^{(\cdot)}}{\alpha+\phi G(\tau_c;\theta)}\right)$. The predictive distribution for $N^+_c$ conditional on the random effect is: 
\begin{equation}
N^+_c|\lambda^o_c,\theta\sim \mbox{Pois}\left(\lambda^o_c\int_{t^+-1}^{t^+}g(s;\theta)\;\dd s\right) = \mbox{Pois}\left(\lambda^o_c G^+_\theta\right),\label{eqn:pred_path}
\end{equation}
where $G^+_\theta = \int_{t^+-1}^{t^+}g(s;\theta)\;\dd s$.

Marginalising over the random effect posterior, we arrive at the negative binomial distribution:
\begin{equation}
\mathbb{P}(N^+_c=n|\alpha^*_c,\phi^*_c) = \frac{\Gamma(\alpha_c^*+n)}{\Gamma(\alpha_c^*)n!}\left(\frac{\alpha_c^*}{\alpha_c^*+\phi^*_cG^+_\theta}\right)^{\alpha^*_c}\left(\frac{\phi^*_c G^+_\theta}{\alpha_c^*+\phi^*_cG^+_\theta}\right)^{n},\quad n\in\mathbb{N}.\label{eqn:pred}
\end{equation}
The length of interval to $t^+$ does not need to be a day and could instead be a week or a month, depending on the context of the application. To obtain the full marginal predictive, we sample the recruitments conditional on parameters sampled from the posterior. For as yet unopened centres, we set $n_c^{(\cdot)}=\tau_c=0$. For each triplet (or couplet, if $\kappa=0$) of parameters sampled from the posterior, we sample $N_c^+$, $c=1,\dots,C$, and sum them to obtain a sample from $N^+|\alpha,\phi,\theta$. The collection of these sums is a sample from the posterior predictive distribution for the model.

If simulations for multiple distinct time periods are required for a given centre, $c$, as needed for the accrual curve for example, then we first sample $\lambda^o_c$ from its posterior \eqref{eqn:re-est}. We then simulate the Poisson counts for the individual time periods, which are conditionally independent given $\lambda_c^o$, from \eqref{eqn:pred_path}.

\subsection{Model averaging}\label{sec:bma}
When predicting the enrolments using a fitted model, we implicitly assume that a single model best reflects reality; however, prediction methods should consider the uncertainty in the models used for inference. We shall, therefore, use model averaging for making predictions, that is, take a weighted average of predictions made by each model. Working in the Bayesian paradigm provides us with an intuitive choice for weights in the form of marginal likelihoods of the models.
\begin{align*}
\Pr(N^+ = n^+|\mathbf{n},\boldsymbol\tau) = \sum_{k=1}^{K}\Pr(N^+ = n^+|\mathbf{n},\boldsymbol\tau, M_k)\pi_p(M_k|\mathbf{n},\boldsymbol\tau),\label{eqn:bma}
\end{align*} 
where $\pi_p(M_k|\mathbf{n},\boldsymbol\tau) \propto \pi(\mathbf{n}|\boldsymbol\tau,M_k)\pi_0(M_k)$, $k=1,\ldots,K$,
with $\pi_0(M_k)$ being prior model probabilities. The averaging framework fits in with the restriction of the shape parameter $\kappa$  to a discrete space. Each $\kappa$ value generates an inhomogeneous Poisson-gamma model with the tail behaviour of the associated intensity shape. This includes the null ($\kappa=0$) model as in \cite{AnFe2007}. In this work we set all prior model probabilities equal.

\subsection{Model validation}
Before making any statements in regards to the future recruitments, we should validate that the fitted model does indeed capture the true data-generating process sufficiently well. Since the true process is unknown, we compare the observed data to the modal model (the model with the highest posterior probability) fixed at posterior parameter means $(\hat{\alpha},\hat{\phi},\hat{\theta})$. 

Firstly, we wish to assess that the chosen hierarchical structure is reflected in the data. The distribution of posterior means of the individual random effects should approximately follow the hierarchical $\mbox{Gamma}(\hat{\alpha},\hat{\alpha}/\hat{\phi})$ distribution. A QQ-plot can be used to visually compare the distributions. If deemed sufficiently similar, using the distribution for generating predictions for yet-unopened centres is appropriate. If the distributions are noticeably different, particularly if the true distribution is multimodal, any interim predictions for yet-unopened centres could (but need not; see robustness study in Section \ref{sec:sim}) be inaccurate.

According to the model, the counts in any initial period $[0,t']$ (such as the first month) of each centre's recruitment period, follow a negative binomial distribution with shape parameter $\alpha$ and success probability $\phi G(t';\theta)/(\alpha+\phi G(t';\theta))$, similar to that given in \eqref{eqn:pred} but using $\alpha$ and $\phi$ in place of $\alpha_c^*$ and $\phi_c^*$. As the true parameters are unknown, we compare it to the distribution fixed at point-estimates $(\hat{\alpha},\hat{\phi},\hat{\theta})$. The diagnostic indicates if the combination of the gamma random effects and the modal decay model captures the behaviour over the initial period after centre initiation.  Again, a QQ-plot can be used for comparing the theoretical distribution to the observation, giving an indication if the fitted model under- or overestimates initial recruitment. The initial period, $[0,t']$, should be long enough that the true recruitment decay should be apparent. However, since only centres that have been recruiting for a period of at least $t'$ can be used for the diagnostic, to ensure a reasonable power, $t'$ should be short enough that a large number of sites have been recruiting for this duration. In this work, we set $t' = 60$ (2 months).

\section{Simulation results}\label{sec:sim}
We demonstrate our flexible framework through a simulation study, using simulated data sets to illustrate model fit and prediction and to highlight the effect model misspecification can have on predictions. In practice, patterns in centre initiation times can  vary greatly between trials. For presenting the methodology, we consider an initiation schedule similar to that observed in a typical trial. We test the robustness of the method using a uniform initiation schedule, with another type of schedule examined in Appendix G of the Supplementary Material. 

Our historical data set do not include the initiation times of the centres, so instead, to accurately reflect the historical data used in the meta-analysis and what is often available to researchers, we take the first recruitment time of a centre as its initiation time and adjust the models to include a single deterministic recruitment at the initiation time of each centre followed by stochastic recruitment as described in Section \ref{sec:model}.

We simulate a study over a course of $600$ days, with $200$ centres. The parameters used for simulations were $\alpha = 1.4$, $\phi = 0.01$, $\kappa = 2.7$ and $\theta = 0.02$. The inference is carried out on data observed in the first $360$ days. As motivated in Section \ref{sec:intro},  we condition the inference on a set of known initiation times, chosen by the practitioner; these could subsequently be varied to investigate the impact of different schedules or initiation models. We consider a set of models with flexible tails (Section \ref{sec:int}) allowing $\kappa\in\{0,0.5,1,2,\infty\}$, thus including the null model \citep{AnFe2007}. The ``normalisation'' of the curve-shapes  was imposed at $\bar{\tau}=\frac{1}{C}\sum_{c=1}^{C}\tau_c$. We purposely simulated using a $\kappa$ value outside of those considered in our models to illustrate the flexibility of the framework. For Bayesian inference, we used parameter and model priors outlined in Sections \ref{sec:prior} and \ref{sec:bma} respectively. Based on the model fitted to the data at the census day $360$, we wish to predict the daily accrual until day $600$.

\begin{figure}
\centering
\includegraphics[width=0.8\textwidth]{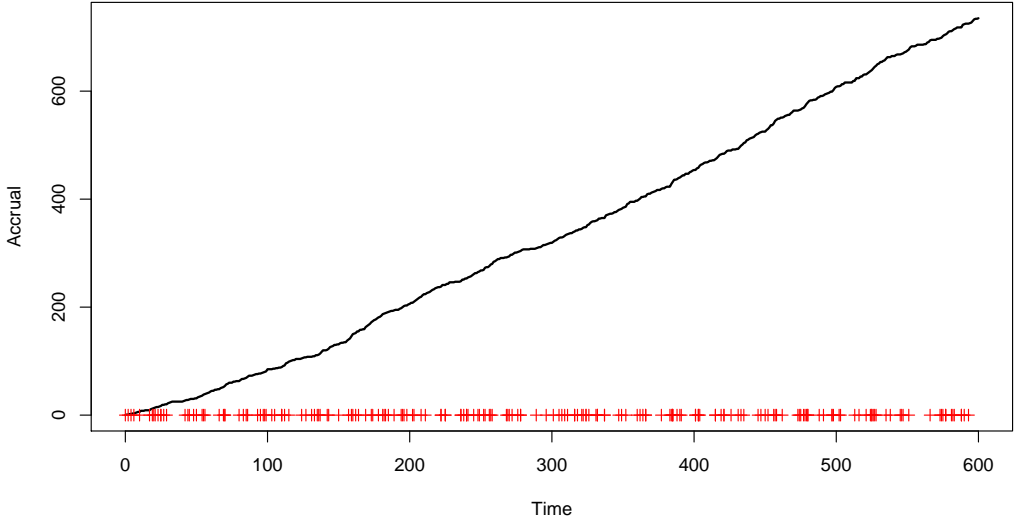}
\caption{Accrual plot with the centre opening times marked by ``+'' symbols on the abscissa.}\label{fig:accrual}
\end{figure}

\begin{table}
\centering

\begin{tabular}{|c|c|c|c|c|c|}
\hline
$\kappa$&$\alpha$&$\phi$&$\theta$&$\pi(\Mb_k|\bm{n})$&ESS\\
\hline
$0$&$1.141\;(0.771, 1.672 )$&$0.013\;(0.011, 0.017)$&$--$&$3.49\times 10^{-25}$&9006\\
$0.5$&$1.167\;(0.759, 1.745)$&$0.013\;(0.010, 0.016)$&$0.143\;(0.044, 0.441)$&$5.51\times 10^{-4}$&8519\\
$1$&$1.144\;(0.742, 1.744)$&$0.013\;(0.011, 0.016)$&$0.033\;(0.021, 0.049 )$&$2.21\times 10^{-1}$&8665\\
$2$&$1.142\;(0.728, 1.644)$&$0.014\;(0.011, 0.016)$&$ 0.017\;(0.012, 0.023 )$&$6.58\times 10^{-1}$&8564\\
$\infty$&$1.122\;(0.718, 1.645)$&$0.014\;(0.011, 0.017)$&$ 0.009\;(0.007, 0.011  )$&$1.20\times 10^{-1}$&8610\\
\hline
\end{tabular}
\caption{Posterior means and 95\% credible intervals, posterior model probabilities and effective sample sizes, obtained using $10^4$ importance samples for each model. }\label{tab:post}
\end{table}

Performing the LRT and BST from Section \ref{sec:detect}, we find the $p$-values of both tests to be $<0.001$. Table \ref{tab:post} provides the fits for the five models. The effective samples sizes are high, which means that each of the model posteriors is represented well by its respective sample and that the marginal likelihood estimates are accurate. If the ESS values had been low, we would have retried using more samples in the importance sampler. We see that model corresponding to $\kappa=\infty$ has the highest posterior probability. A trellis plot of the posteriors for $(\tilde{\alpha},\tilde{\phi},\tilde{\theta})$ from the modal model (see Appendix G of the Supplementary Material)  confirms at least approximate pairwise orthogonality between the parameters, as anticipated from Sections \ref{sec:model} and \ref{sec:prior}. QQ-plots for the modal model comparing the hierarchical gamma distribution to the posterior means of the random effects, and comparing the observed recruitments over the first two months of each centre's recruiting period to the model's negative binomial distribution both show approximate straight lines with unit gradient and are provided in the Supplementary Material. 

Figure \ref{fig:sim_fcast} shows the accrual forecast from the census time $\tau = 360$ up to the horizon $\tau^H=600$, superimposed onto the true accrual plot. The forecast is based on the Bayesian model-averaged posterior predictive distribution. The true accrual is contained within the 95\% predictive intervals. 

\begin{figure}
\begin{subfigure}{.5\textwidth}
  \centering
  \includegraphics[width=.9\textwidth]{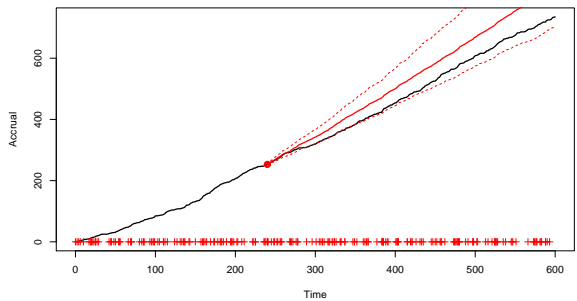}
  \caption{\normalsize{ Bayesian model averaging}}
  \label{fig:sim_early_bma}
\end{subfigure}%
\begin{subfigure}{.5\textwidth}
  \centering
  \includegraphics[width=.9\textwidth]{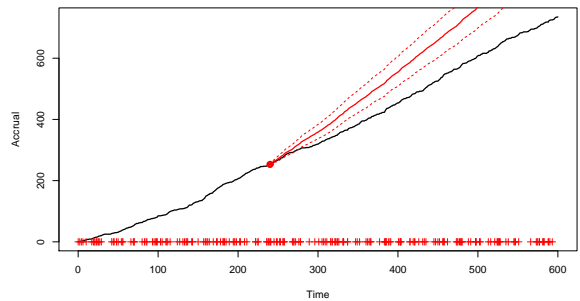}
  \caption{\normalsize{ Maximum likelihood and model selection}}
  \label{fig:sim_early_mle}
\end{subfigure}%
\caption{Comparison of accrual predictions produced by two methods; accruals (black, solid) with predictive means (red, solid) and 95\% prediciton bands (red, dashed).  Prediction bands are based on the $2.5\%$ and $97.5\%$ quantiles. The ``+'' symbols on the abscissa indicate centre opening times.}
\label{fig:sim_mle_comp}
\end{figure}

Figures \ref{fig:sim_early_bma} and \ref{fig:sim_early_mle} use an earlier census time ($\tau=240$) to  illustrate the issues that can arise when making predictions  using maximum likelihood estimation and model selection. The inference was carried out with the same set of candidate models, and predictions were obtained by simulating from the best model ($\kappa=\infty$, chosen using AIC) with parameters fixed at the MLEs. As shown in the plots, not accounting for parameter and model uncertainty may lead to overly confident and biased predictions. Simulations with $\tau=360$ (see Supplementary material) still showed bias due to the choice of a single model, although the contrast with Figure \ref{fig:sim_fcast} in terms of prediction interval width was less marked.

\begin{figure}
\centering
\includegraphics[width=0.8\textwidth]{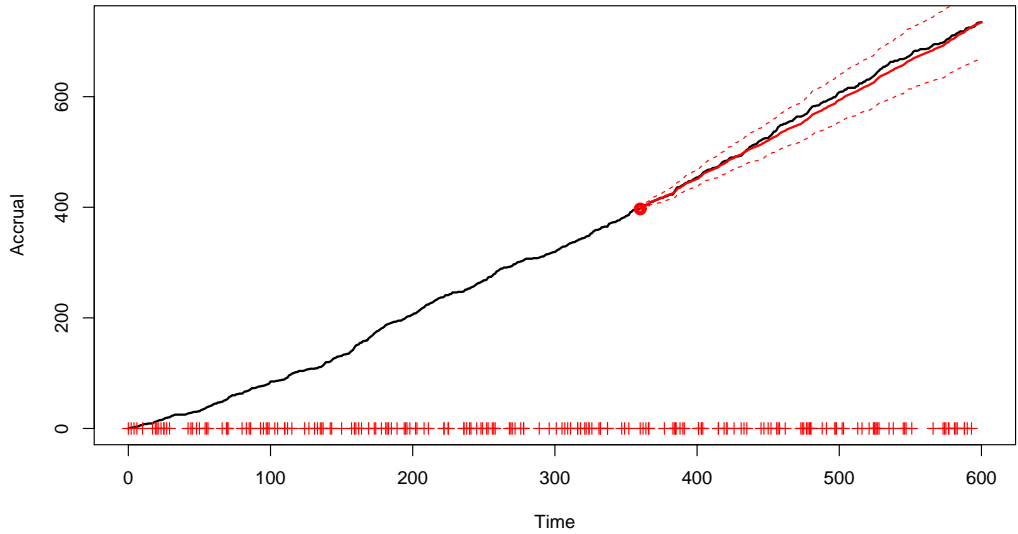}
\caption{Accrual with Bayesian model-averaged forecast predictive mean (solid, red) and 95\% prediction bands (red, dashed). Prediction bands are based on the $2.5\%$ and $97.5\%$ quantiles. The forecast begins from a point marked by the red dot and the ``+'' symbols on the abscissa indicate centre opening times.} 
\label{fig:sim_fcast}
\end{figure}

\begin{figure}
\centering
\includegraphics[width = 0.8\textwidth]{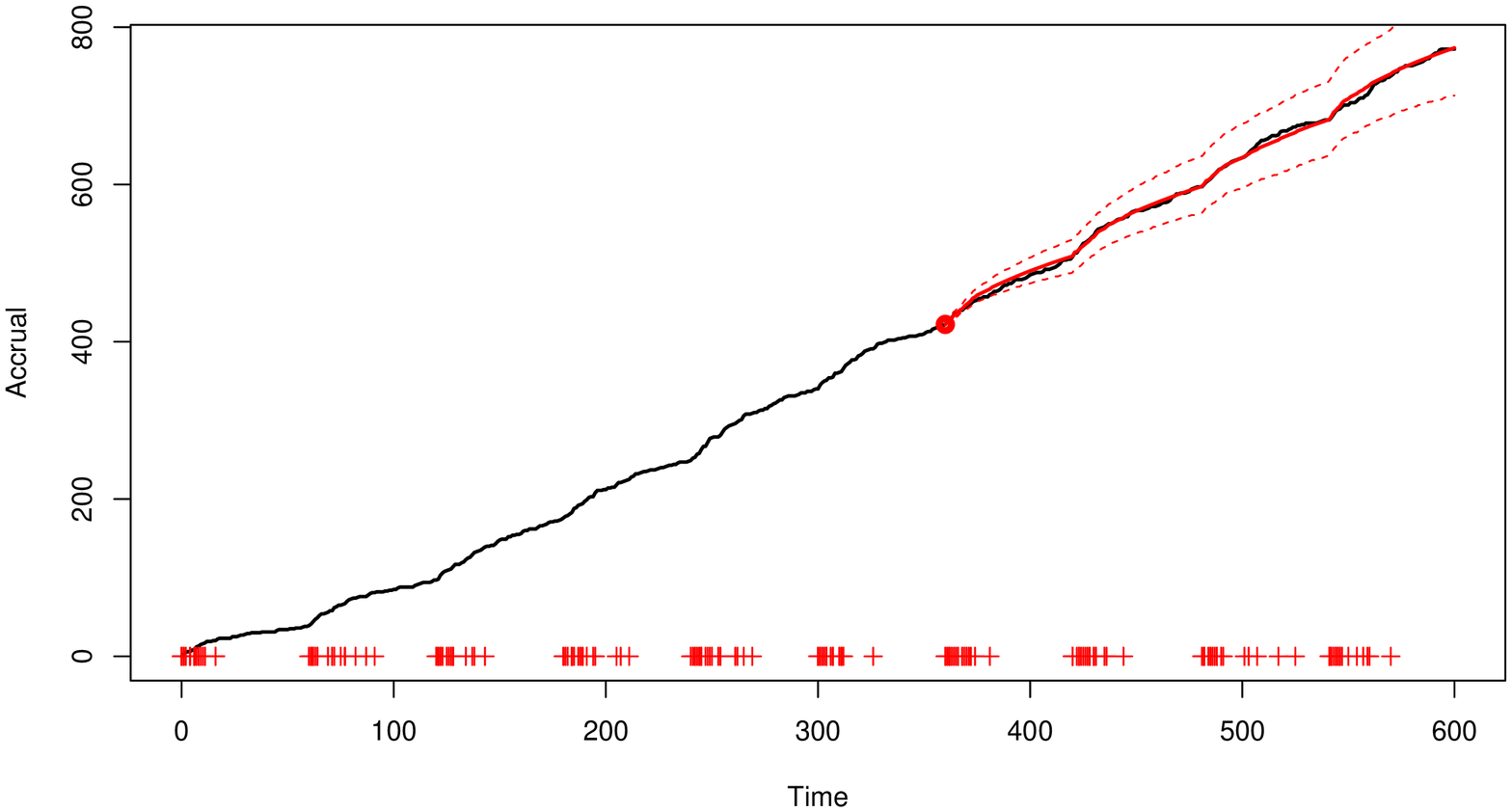}
\caption{Accrual with forecast predictive mean (solid, red) and 95\% prediction bands (red, dashed).  Prediction bands are based on the $2.5\%$ and $97.5\%$ quantiles. The forecast begins from a point marked by the red dot and the ``+'' symbols on the abscissa indicate centre opening times.} 
\label{fig:sim_clump}
\end{figure}

We repeated the analysis with a different distribution of initiation times, making the centre initiations ``clump'' roughly every two months. The resulting forecast predictive distribution can be seen in Figure \ref{fig:sim_clump}; performance appears to be robust to the type of initiation schedule. 

\begin{figure}
\begin{subfigure}{.5\textwidth}
  \centering
  \includegraphics[width=.9\textwidth]{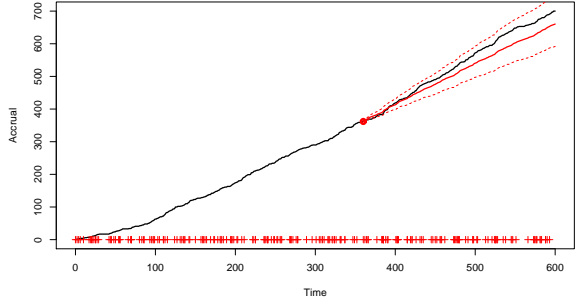}
  \caption{\normalsize{ Uniform openings, $\mathbb{E}[\lambda^o_c] = 0.01$;\\
  $p$-value = 0.212}}
  \label{fig:sim_bim1}
\end{subfigure}%
\begin{subfigure}{.5\textwidth}
  \centering
  \includegraphics[width=.9\textwidth]{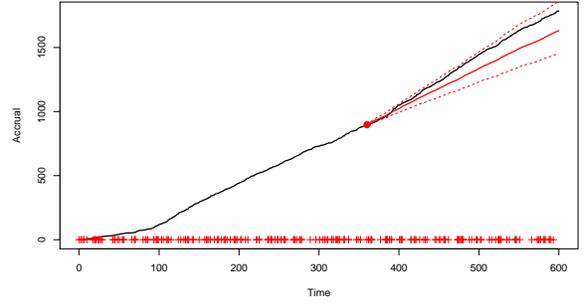}
  \caption{\normalsize{ Uniform openings, $\mathbb{E}[\lambda^o_c] = 0.03$;\\
  $p$-value = 0.207}}
  \label{fig:sim_bim3}
\end{subfigure}%
\caption{Accruals (black, solid) with predictive means (red, solid) and 95\% prediciton bands (red, dashed) when the true random-effect distribution is a mixture.  Prediction bands are based on the $2.5\%$ and $97.5\%$ quantiles. The ``+'' symbols on the abscissa indicate centre opening times.}
\label{fig:sim_bim}
\end{figure}

To further test the robustness of the framework, we first consider the random effects $\lambda^o_c$ now being generated from a mixture of two gamma distributions
\begin{equation*}
\lambda^o_c|\alpha,\phi_1,\phi_2\sim \frac{1}{2}\mbox{Gamma}\left(\alpha,\frac{\alpha}{\phi_1}\right)+\frac{1}{2}\mbox{Gamma}\left(\alpha,\frac{\alpha}{\phi_2}\right).\label{eqn:mix}
\end{equation*}
We considered data generated using the same $\alpha$ value and curve-shape as before, but now with centre initiation times uniformly sampled on the interval. The ratio of gamma expectations was fixed such that $\phi_2 = 10\phi_1$, and the random effect expectation, $E[\lambda^o_c] = (\phi_1+\phi_2)/2$, was set to 0.01 and then 0.03. Figures \ref{fig:sim_bim1} and \ref{fig:sim_bim3} show example forecasts for accruals with the two different expectations. The more data, that is, the larger $E[\lambda_c^o]$, the more apparent the discrepancy in the random-effect distribution, and the concomitant predictions, becomes.  This is visible in the clearly non-linear  diagnostic QQ-plots, and the plotted forecasts (see Supplementary Material). 
The robustness of predictions comes from the fact that the random effects for initiated centres use re-estimated data-driven distributions, reducing the importance of the random-effect prior; thus the main source of forecasting error comes from the incorrect random-effect prior for new centres. Similar plots for the "clumped" initiation schedule, provided in the Supplementary material, show the same pattern.
This mixture distribution of random effects represents the (extreme) scenario where roughly half of the centres recruit the vast majority of patients, with the remaining sites recruiting little to none each. When the ratio of the two means is closer to 1, the model still produces reliable predictions.
 
\begin{figure}
\begin{subfigure}{.5\textwidth}
  \centering
  \includegraphics[width=.9\textwidth]{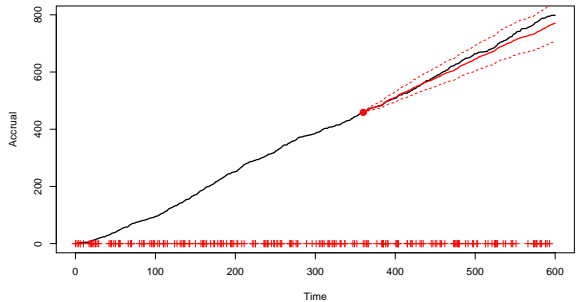}
  \caption{\normalsize{Uniform openings, $\mathbb{E}[\lambda^o_c] = 0.01$;\\
  $p$-value = 0.553}}
  \label{fig:sim_weib1}
\end{subfigure}%
\begin{subfigure}{.5\textwidth}
  \centering
  \includegraphics[width=.9\textwidth]{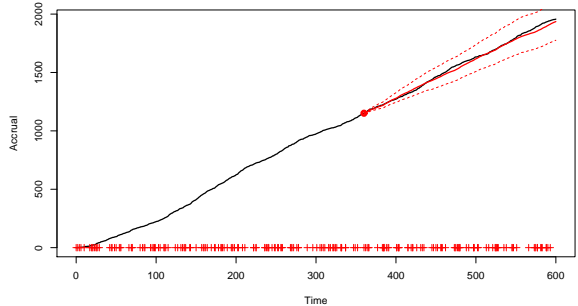}
  \caption{\normalsize{Uniform openings, $\mathbb{E}[\lambda^o_c] = 0.03$;\\
  $p$-value = 0.033}}
  \label{fig:sim_weib3}
\end{subfigure}%
\caption{Accruals (black, solid) with predictive means (red, solid) and 95\% prediciton bands (red, dashed) when the true intensity shape is Weibull, for two different values of $\mathbb{E}[\lambda_c^o]$.  Prediction bands are based on the $2.5\%$ and $97.5\%$ quantiles. The ``+'' symbols on the abscissa indicate centre opening times.}
\label{fig:sim_weib}
\end{figure}

We also consider the effect of curve-shape misspecification on predictions, generating data using an intensity proportional to the Weibull density function
\begin{align*}
g_W(t;\theta,k) &= \frac{\frac{k}{\theta}\left(\frac{t}{\theta}\right)^{k-1}\exp\{-(t/\theta)^{k}\}}{1-\exp\{-(\tau/\theta )^{k}\}}\tau,
~~~\mbox{so}~~~
G_W(t;\theta,k) = \frac{1-\exp\{-(t/\theta )^{k}\}}{1-\exp\{-(\tau/\theta )^{k}\}}\tau,
\end{align*}
where $\theta$, $k>0$. We simulated accrual datasets using the Weibull shape with $\theta=30$ and $k=1.5$, resulting in the highest recruitment rates occurring two weeks after centre initiation. The random-effect distribution used $\alpha=1.4$ and two different values $\phi$ were used:  $0.01$ and $0.03$;  Figures \ref{fig:sim_weib1} and \ref{fig:sim_weib3} show example forecasts. For lower overall recruitment levels, the model still  predicts future accrual well. Forecast inaccuracies due to model misspecifiation become more apparent when larger recruitment rates are used. The same pattern is observed when centre initiation times are clumped (see Appendix G of the Supplementary Material).

\section{Data results}\label{sec:data}
\begin{figure}
    \centering
    \begin{minipage}{0.45\textwidth}
        \centering
        \includegraphics[width = 0.8\textwidth]{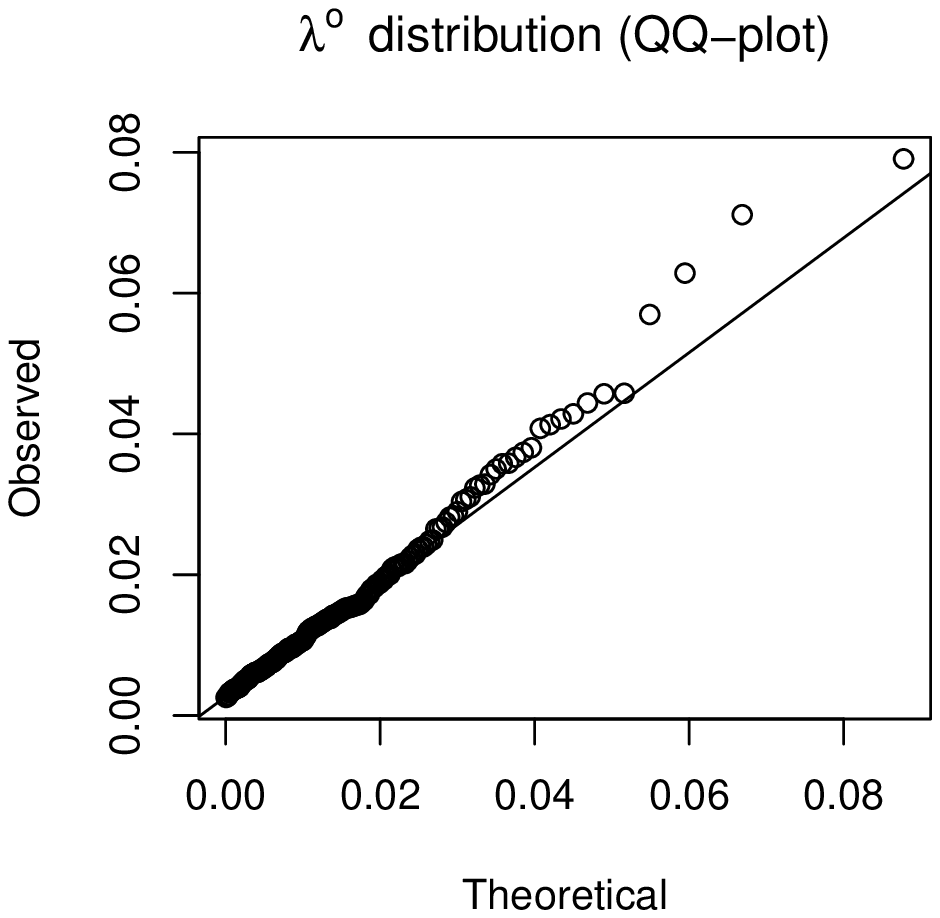}
\caption{Re-estimated $\lambda^o_c$ expectations compared to Gamma$\left(\hat\alpha,\hat{\alpha}/\hat{\phi}\right)$ distribution.}
\label{fig:data_diag1}
    \end{minipage}\hfill
    \begin{minipage}{0.45\textwidth}
        \centering
        \includegraphics[width = 0.8\textwidth]{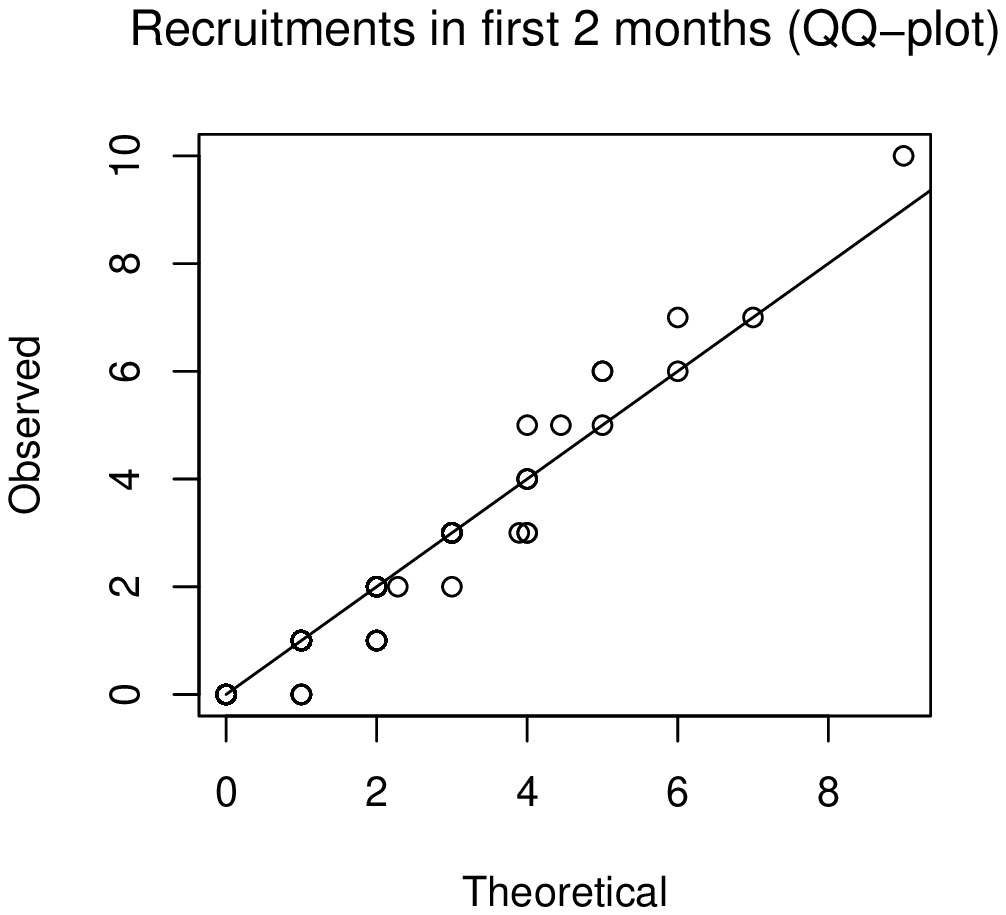}
\caption{Observed recruitments compared to the theoretical negative binomial distribution.}
\label{fig:data_diag2}
    \end{minipage}
\end{figure}

\begin{figure}
\centering

\includegraphics[width=0.8\textwidth]{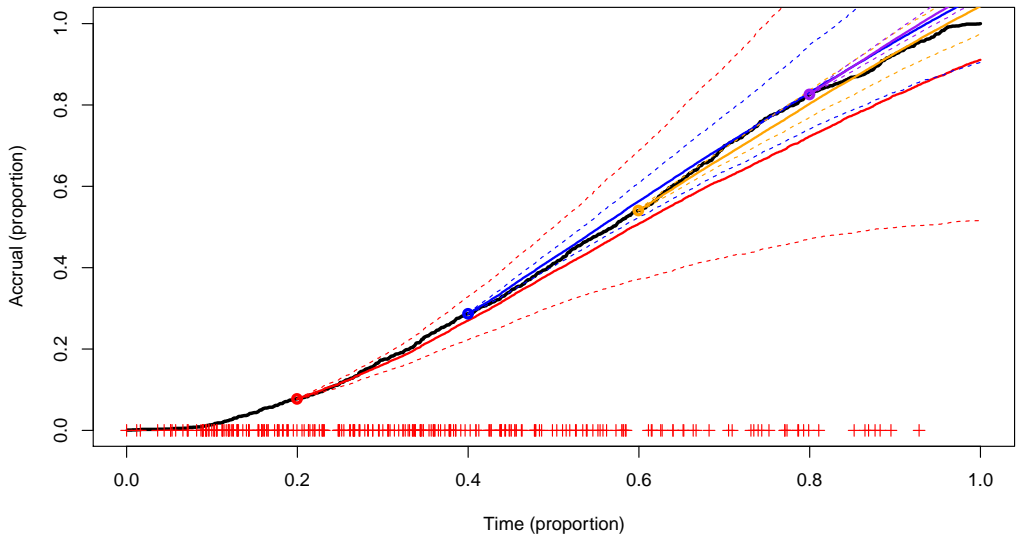}
\caption{Accrual (black, solid) for an oncology study; coloured solid lines are mean predictions from census times, dashed lines are the $95\%$ prediction bands, and the ``+'' symbols indicate opening times of centres.}\label{fig:data}
\end{figure}

We fitted the same set of models to a recruitment dataset of a prostate-cancer clinical trial. The recruitment was carried out across 244 sites. The accrual is presented as the proportion of the total number enrolled. Similarly, time is given as the proportion of the total recruiting period. Figures \ref{fig:data_diag1} and \ref{fig:data_diag2} show the diagnostic QQ-plots for the model fitted to data available at time $0.4$. They indicate that there is sufficient concordance between the assumed model and observed enrolment giving validity to potential predictions. Figure \ref{fig:data} shows the accrual along with forecasts from four different census times.  The predictive bands become narrower and parameter uncertainty decreases at each census as more data become available for inference. After the third census, there is an unexpected jump in accrual followed by a drop around the fourth census time, suggesting a global external factor, such as a change in the protocol.  Table \ref{tab:data} shows $p$-values of the LRT and BST. Initially, when the accrual is still only a small proportion of the total, it is hard to detect the time-inhomogeneity. At later census points, the test outcomes indicate that the rates are not constant.

\begin{figure}
\centering
\includegraphics[width=0.8\textwidth]{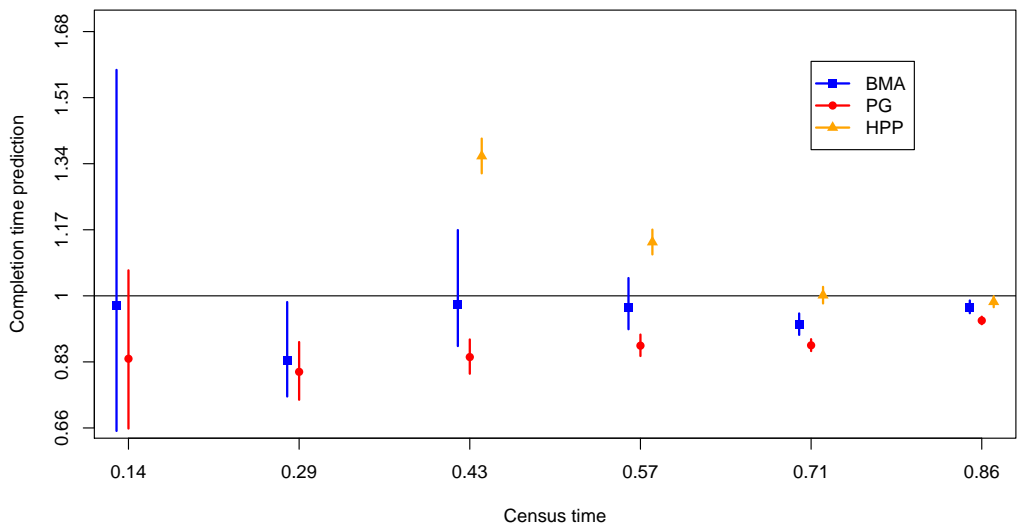}
\caption{Predictive distributions for time needed to make the final recruitment  in the data example in Section 7, as forecast by three different modelling frameworks: Bayesian model averaging (BMA), time-homogeneous Poisson-gamma (PG) and homogeneous Poisson process fit to accrual only (HPP). The horizontal line represents the true completion time and the prediction positions of the $x$-axis were off-set by $0.01$ for clarity.}\label{fig:comparison}
\end{figure}

We compare the proposed framework to the standard homogeneous PG model \eqref{eqn:anis} as well as a homogeneous Poisson process (HPP) model fitted only to the accrual. We used the same priors as outlined in Section \ref{sec:prior} for fitting the PG model, and the HPP rate estimate was obtained using maximum likelihood. The methods were compared in terms of the predicted completion time of the recruitment for the study with the sampling details outlined in Appendix F of the Supplementary Material. Forecast completion time from 6 different census points and can be seen in Figure \ref{fig:comparison}; the first HPP predictions were centred at $3.67$ and $1.84$ which were outside the plot's range. The proposed framework produces better point predictions, especially at earlier interim analyses, and more closely represents the true uncertainty. The HPP predictions near the end of the trial are very accurate. At this point, the majority of the centres having already been initiated and have been recruiting for a long period of time. As a result, the total recruitment rates are not changing by much, with the slight decreasing trend offset by the occasional initiation of a new centre. This is a coincidence; if the decay rate had been sharper or shallower, or if fewer or more centres had been initiated then the naive overall Poisson process model would not have fitted as well. The underprediction of the completion time by the proposed model at the census time of $t=0.71$ is likely a result of the unexpected surge in recruitment at around that time. The surge is examined in more detail in Appendix G of the Supplementary Material.

\begin{table}
\centering
\begin{tabular}{|c|c|c|c|}
\hline
Census time&BST $p$-value&LRT $p$-value&Forecast $p$-value\\
\hline
1&0.196&0.226&0.697\\
2&0.012&0.021&0.625\\
3&$<0.001$&$<0.001$&0.029\\
4&$<0.001$&$<0.001$&$<0.001$\\
\hline
\end{tabular}
\caption{Decay in rate test $p$-values and the forecasting $p$-values at four census times. }\label{tab:data}
\end{table}

\section{Discussion}\label{sec:disc}

We have introduced a general, flexible framework for modelling and predicting recruitment to clinical trials. We suggest two tests for detecting decay in recruitment rates; comparing them both with respect to power and robustness. The particular form of the test statistic allows for a single, simple trial-level test. Alternative forms, such as splitting according to a global time, would either require a test for each centre, massively reducing the power, or estimates of all of the individual centre intensities which would introduces several layers of additional complexity because of the hierarchical connection between the centre intensities. If it were believed \emph{a priori} that a particular global period would be unrepresentative then this time span, and the concomitant recruitment, could simply be removed, albeit at the cost of lower power.

The parametric curve-shape forms chosen for the intensity were based on the features encountered in oncology trials. We found that the model was still robust to moderate model misspecifications in the distribution of the random effect and intensity shape. Other therapeutic areas such as pulmonary or cardio-vascular diseases experience more frequent recruitments and different curve-shapes may be appropriate. As shown in Section \ref{sec:sim}, model misspecification becomes more of a problem at larger enrollment rates. However, with increased frequency, pattern changes in the early months of a centre are easier to identify. Using more complex parametric forms, such as Weibull or generalised gamma shape, could lead to more accurate predictions. Alternatively, if covariate information is available, say $\mathbf{x}_c$ for each centre, the following intensity form motivated by hazard models from survival analysis could be used: $
\lambda_c(t) = \lambda_c^o\exp\{\beta^\top \mathbf{x}_c\}g\left(t;\exp\{\eta^\top \mathbf{x}_c\}\right),$ where $\lambda^o_c$ are now random effects coming from a $\mbox{Gamma}(\alpha,\alpha)$ distribution and $\beta$ and $\eta$ are vectors of unknown parameters.

As seen in the data example in Section \ref{sec:data} there can be external factors modulating the overall accrual. This could potentially be modelled via a short-term, constant global intensity modifier, which would maintain tractability.  The framework is not constrained to parametric forms; non-parametric intensity models, such as those using B-splines (for example, \cite{MoNe2019}) or Gaussian processes (for example, \cite{AdMu2009}), could be used instead. This, however, would make the intensity extrapolation problem more difficult.

For curve-shape parameter prior construction, our choice of the quantity of interest $R_\kappa$ was motivated by simplicity of the form; one could just as well have used $\frac{G_\kappa(t_0/2;\theta)}{G_\kappa(t_0;\theta)}$, albeit with more algebraic manipulations. The general method was aimed at models with monotonically decreasing intensities. If curve-shapes such as Weibull are considered then constructing sensible priors will be  more complicated.

In presenting the method, we condition the inference and prediction on known initiation schedules for the centres. Incorporating stochastic centre initiation models, such as those in \cite{Anis2009} and \cite{LaTa2018}, into the Monte Carlo prediction framework is  straight-forward, but would complicate the presentation of our methodology without adding novelty. In Appendix H of the Supplementary Material, we demonstrate how recruitment can be predicted using our methodology when there is uncertainty in the initiation schedule. For illustration, we imagine a Weibull-distributed delay to each centre's initiation, but any other initiation model could be incorporated in a similar manner. We stress that full prediction intervals should take this uncertainty into account.

In this work, we focus on patient recruitment regardless of the numbers of dropouts observed. In practice,  screening failure and patient withdrawal are both prevalent in clinical trials. Assuming the dropouts are independent of the recruitment process, existing survival analysis techniques such as Cox's proportional hazard model \citep{CoxD1972} or accelerated failure time frailty model \citep{WeiL1992} could be used in combination with the recruitment model to produce distributions of the numbers of patients in the system at a given time. Such knowledge would be useful to the practitioners and operational researchers in charge of drug-supply chains for the centres.

\cite{AnFe2007} introduced a method for determining the number of additional centres needed to be initiated for the study to finish on time. With minimal adaptation, the same method can also be used with our model. However, since it assumes that all new centres are initiated immediately, it may not apply in all scenarios. We would advocate a simulation-based approach, where forecasts based on different centre initiation schedules are compared. As different operational costs can be associated with different schedules, this would become a resource-constrained optimisation problem.
\section{Software}
Software in the form of \texttt{R} code is available at \url{https://github.com/SzymonUrbas/ct-recuitment-prediction}.
\section*{Acknowledgements}
This work was supported by the Engineering and Physical Sciences Research Council (grant number EP/L015692/1) and AstraZeneca.

\bibliography{draft_ref}

\begin{thebibliography}{}

\bibitem[Adams et~al., 2009]{AdMu2009}
Adams, R.~P., Murray, I., and MacKay, D.~J. (2009).
\newblock Tractable nonparametric {B}ayesian inference in {P}oisson processes
  with {G}aussian process intensities.
\newblock In {\em Proceedings of the 26th Annual International Conference on
  Machine Learning}, pages 9--16.

\bibitem[Akaike, 1973]{Akai1973}
Akaike, H. (1973).
\newblock Information theory and an extension of the maximum likelihood
  principle.
\newblock Petrov BN, Csaki F, editors. Second International Symposium on
  Information Theory, pages 267--281. Budapest (Hungary): Akademiai Kiado.

\bibitem[Anisimov, 2009]{Anis2009}
Anisimov, V. (2009).
\newblock Predictive modelling of recruitment and drug supply in multicenter
  clinical trials.
\newblock In {\em Proc. of Joint Statistical Meeting}, pages 1248--1259.

\bibitem[Anisimov and Fedorov, 2007]{AnFe2007}
Anisimov, V.~V. and Fedorov, V.~V. (2007).
\newblock Modelling, prediction and adaptive adjustment of recruitment in
  multicentre trials.
\newblock {\em Statistics in Medicine}, 26(27):4958--4975.

\bibitem[Cox, 1972]{CoxD1972}
Cox, D.~R. (1972).
\newblock Regression models and life-tables.
\newblock {\em Journal of the Royal Statistical Society: Series B
  (Methodological)}, 34(2):187--202.

\bibitem[Devroye, 1986]{Devr1986}
Devroye, L. (1986).
\newblock {\em Non-uniform Random Variate Generation}.
\newblock New York: Springer-Verlag.

\bibitem[Doucet et~al., 2013]{Smit2013}
Doucet, A., Freitas, N.~d., and Gordon, N. (2013).
\newblock {\em Sequential Monte Carlo methods in practice}.
\newblock New York: Springer Science.

\bibitem[Gajewski et~al., 2008]{GaSi2008}
Gajewski, B.~J., Simon, S.~D., and Carlson, S.~E. (2008).
\newblock Predicting accrual in clinical trials with {B}ayesian posterior
  predictive distributions.
\newblock {\em Statistics in Medicine}, 27(13):2328--2340.

\bibitem[Gelman et~al., 2013]{GeCa2013}
Gelman, A., Carlin, J.~B., Stern, H.~S., Dunson, D.~B., Vehtari, A., and Rubin,
  D.~B. (2013).
\newblock {\em Bayesian data analysis}.
\newblock Chapman and Hall/CRC.

\bibitem[Getz and Lamberti, 2013]{Tuft2013}
Getz, K. and Lamberti, M.~J. (2013).
\newblock 89\% of trials meet enrolment, but timelines slip, half of sites
  under-enrol.
\newblock {\em Tufts {CSDD} Impact Report}, 15:1--4.

\bibitem[Gu et~al., 2008]{GuNg2008}
Gu, K., Ng, H. K.~T., Tang, M.~L., and Schucany, W.~R. (2008).
\newblock Testing the ratio of two {P}oisson rates.
\newblock {\em Biometrical Journal: Journal of Mathematical Methods in
  Biosciences}, 50(2):283--298.

\bibitem[Heitjan et~al., 2015]{HeGe2015}
Heitjan, D.~F., Ge, Z., and Ying, G.~S. (2015).
\newblock Real-time prediction of clinical trial enrollment and event counts: a
  review.
\newblock {\em Contemporary Clinical Trials}, 45:26--33.

\bibitem[Hjort and Claeskens, 2003]{HjCl2003}
Hjort, N.~L. and Claeskens, G. (2003).
\newblock Frequentist model average estimators.
\newblock {\em Journal of the American Statistical Association},
  98(464):879--899.

\bibitem[Huzurbazar, 1950]{Huzu1950}
Huzurbazar, V.~S. (1950).
\newblock Probability distributions and orthogonal parameters.
\newblock volume~46 of {\em Mathematical Proceedings of the Cambridge
  Philosophical Society}, pages 281--284. Cambridge University Press.

\bibitem[Jiang et~al., 2015]{JiSi2015}
Jiang, Y., Simon, S., Mayo, M.~S., and Gajewski, B.~J. (2015).
\newblock Modeling and validating {B}ayesian accrual models on clinical data
  and simulations using adaptive priors.
\newblock {\em Statistics in Medicine}, 34(4):613--629.

\bibitem[Kasenda et~al., 2014]{KaEl2014}
Kasenda, B., Von~Elm, E., You, J., Bl{\"u}mle, A., Tomonaga, Y., Saccilotto,
  R., Amstutz, A., Bengough, T., Meerpohl, J.~J., Stegert, M., et~al. (2014).
\newblock Prevalence, characteristics, and publication of discontinued
  randomized trials.
\newblock {\em JAMA}, 311(10):1045--1052.

\bibitem[Krishnamoorthy and Thomson, 2004]{KrTh2004}
Krishnamoorthy, K. and Thomson, J. (2004).
\newblock A more powerful test for comparing two {P}oisson means.
\newblock {\em Journal of Statistical Planning and Inference}, 119(1):23--35.

\bibitem[Lan et~al., 2019]{LaTa2018}
Lan, Y., Tang, G., and Heitjan, D.~F. (2019).
\newblock Statistical modeling and prediction of clinical trial recruitment.
\newblock {\em Statistics in Medicine}, 38:945--955.

\bibitem[Lasagna, 1979]{Lasa1979}
Lasagna, L. (1979).
\newblock Problems in publication of clinical trial methodology.
\newblock {\em Clinical Pharmacology \& Therapeutics}, 25(5part2):751--753.

\bibitem[Lee, 1983]{Lee1983}
Lee, Y.~J. (1983).
\newblock Interim recruitment goals in clinical trials.
\newblock {\em Journal of Chronic Diseases}, 36(5):379--389.

\bibitem[Lindley, 1957]{Lind1957}
Lindley, D.~V. (1957).
\newblock A statistical paradox.
\newblock {\em Biometrika}, 44(1-2):187--192.

\bibitem[Morgan et~al., 2019]{MoNe2019}
Morgan, L.~E., Nelson, B.~L., Titman, A.~C., and Worthington, D.~J. (2019).
\newblock A spline-based method for modelling and generating a nonhomogeneous
  {P}oisson process.
\newblock In {\em 2019 Winter Simulation Conference (WSC)}, pages 356--367.
  IEEE.

\bibitem[Nelder and Mead, 1965]{NeMe1965}
Nelder, J.~A. and Mead, R. (1965).
\newblock A simplex method for function minimization.
\newblock {\em The computer journal}, 7(4):308--313.

\bibitem[Piantadosi and Patterson, 1987]{PiPa1987}
Piantadosi, S. and Patterson, B. (1987).
\newblock A method for predicting accrual, cost, and paper flow in clinical
  trials.
\newblock {\em Controlled Clinical Trials}, 8(3):202--215.

\bibitem[Robert and Casella, 2013]{RoCa2013}
Robert, C. and Casella, G. (2013).
\newblock {\em Monte Carlo statistical methods}.
\newblock Springer Science \& Business Media.

\bibitem[Robertson et~al., 1988]{RoWr1988}
Robertson, T., Wright, F., and Dykstra, R. (1988).
\newblock {\em Order restricted statistical inference}.
\newblock Wiley: New York.

\bibitem[Schwarz, 1978]{Schw1978}
Schwarz, G. (1978).
\newblock Estimating the dimension of a model.
\newblock {\em The Annals of Statistics}, 6(2):461--464.

\bibitem[Tang et~al., 2012]{TaKo2012}
Tang, G., Kong, Y., Chang, C.-C.~H., Kong, L., and Costantino, J.~P. (2012).
\newblock Prediction of accrual closure date in multi-center clinical trials
  with discrete-time {P}oisson process models.
\newblock {\em Pharmaceutical Statistics}, 11(5):351--356.

\bibitem[Wei, 1992]{WeiL1992}
Wei, L.~J. (1992).
\newblock The accelerated failure time model: a useful alternative to the {C}ox
  regression model in survival analysis.
\newblock {\em Statistics in Medicine}, 11(14-15):1871--1879.

\bibitem[Williford et~al., 1987]{WiBi1987}
Williford, W.~O., Bingham, S.~F., Weiss, D.~G., Collins, J.~F., Rains, K.~T.,
  and Krol, W.~F. (1987).
\newblock The ``constant intake rate'' assumption in interim recruitment goal
  methodology for multicenter clinical trials.
\newblock {\em Journal of Chronic Diseases}, 40(4):297--307.

\bibitem[Zhang and Long, 2010]{ZhLo2010}
Zhang, X. and Long, Q. (2010).
\newblock Stochastic modeling and prediction for accrual in clinical trials.
\newblock {\em Statistics in Medicine}, 29(6):649--658.

\end{thebibliography}
\bibliographystyle{apalike}
\newpage


\appendix
\section*{Supplementary material}
This file contains the technical appendix for ``Interim recruitment prediction for multi-centre clinical trials''. The algorithm for the non-parametric bootstrap test of Section 3 of the main article is outlined in Appendix A. Appendix B provides full parametric forms of the integrated intensity curve-shapes described in Section 4; it also discusses a potential identifiability problem. Appendices C and D outline the details of maximum likelihood and Bayesian inference on the model parameters. Appendix E provides the density of the prior described in Section 5.1. The time-to-completion Monte Carlo sampling algorithm is outlined in Appendix F. Appendix G provides additional details and figures for the simulation study in Section 6 and data analysis in Section 7. Appendix H describes an implementation of a centre-initiation delay model into the prediction framework. 
\section{Non-parametric bootrapped test}\label{app:bst}

\IncMargin{1em}
\begin{algorithm}
\SetKwData{Left}{left}\SetKwData{This}{this}\SetKwData{Up}{up}
\SetKwFunction{Union}{Union}\SetKwFunction{FindCompress}{FindCompress}
\SetKwInOut{Input}{input}\SetKwInOut{Output}{output}

\Input{Series of counts $\{N_c(t)\}_{t=1}^{\tau_c}$, $c = 1,\ldots,C$; number of bootstrapped samples $B$. }
\Output{Probability of observed difference in means under $\Hb_0$.}
\BlankLine
Calculate observed difference $ \Delta = \sum_{c=1}^{C}\left(\sum_{t=1}^{\tau_c/2}N_c(t)- \sum_{t=\tau_c/2+1}^{\tau_c}N_c(t)\right)$\;
\For{$b\leftarrow 1$ \KwTo $B$}{
\For{$c\leftarrow 1$ \KwTo $C$}{
Resample $\{N^{(b)}_c(t)\}_{t=1}^{\tau_c}$ with replacement\;
}
Calculate difference $ \Delta^{(b)} = \sum_{c=1}^{C}\left(\sum_{t=1}^{\tau_c/2}N^{(b)}_c(t)- \sum_{t=\tau_c/2+1}^{\tau_c}N^{(b)}_c(t)\right)$\;
}
Calculate approximate $p$-value: $\hat{p} = \frac{1}{B}\sum_{b=1}^B \mathbb{I}_{\left\{\Delta\geq\Delta^{(b)}\right\}}$
\caption{Non-parametric bootstrapped test}\label{alg:bst}
\end{algorithm}\DecMargin{1em}
\section{Curve-shape}\label{app:int}
The integrated, normalised parametric intensities are:
\begin{align*}
G_0(t) &= t,\\
G_1(t;\theta) &=\frac{\log(1+\theta t)}{\log(1+\theta \tau)}\tau,\\
G_\kappa(t;\theta) &= \frac{(1+\theta t/\kappa)^{1-\kappa}-1}{(1+\theta \tau/\kappa)^{1-\kappa}-1}\tau,\quad \kappa\notin\{0,1,\infty\},\\
G_\infty(t;\theta) &= \frac{1-\exp\{-\theta t\}}{1-\exp\{-\theta\tau\}}\tau.
\end{align*}
In two instances, the flexible-tail form can give rise to identifiability problems:

\underline{$t,\tau>>\kappa/\theta$ and $\kappa<1$}
\begin{align*}
G_\kappa(t;\theta) = \frac{(1+\theta t /\kappa)^{1-\kappa}-1}{(1+\theta \tau /\kappa)^{1-\kappa}-1}\tau\approx \frac{(\theta t /\kappa)^{1-\kappa}-1}{(\theta \tau /\kappa)^{1-\kappa}-1}\tau \approx \left(\frac{t}{\tau}\right)^{1-\kappa}\tau,
\end{align*}
which does not depend on $\theta$.

\underline{$t,\tau>>\kappa/\theta$ and $\kappa>>1$}
\begin{align*}
G_\kappa(t;\theta) &= \frac{(1+\theta t /\kappa)^{1-\kappa}-1}{(1+\theta \tau /\kappa)^{1-\kappa}-1}\tau\\
&\approx\frac{(1+\theta t /\kappa)\exp\{-\theta t\}-1}{(1+\theta \tau /\kappa)\exp\{-\theta \tau\}-1}\tau\\
&\approx \frac{\exp\{-\theta t\}-1}{\exp\{-\theta \tau\}-1}\tau,
\end{align*}
which does not depend on $\kappa$.
\section{Maximum likelihood inference}\label{app:mle}

In the frequentist setting, we aim to find estimators which maximise the likelihood surface (4.2) in the main paper. This is equivalent to maximising the log-likelihood surface (up to a constant)
\begin{align*}
\ell(\alpha,\phi,\theta|\mathbf{n},\boldsymbol\tau) = C\left(\alpha\log\frac{\alpha}{\phi}-\log\Gamma(\alpha)\right)
- \sum_{c=1}^C\left\{\left(\alpha+n^{(\cdot)}_{c}\right)\log\left(G(\tau_c;\theta)+\frac{\alpha}{\phi}\right) \right . &\label{eqn:llik}\\
 \left.-\log\Gamma\left(\alpha+n^{(\cdot)}_{c}\right)- \sum_{t = 1}^{\tau_c} n_c^{(t)}\log( G(t;\theta)-G(t-1;\theta))\right\}.&\nonumber
\end{align*}

The log-likelihood function can be optimised using a range of methods, for example, the Nelder-Mead \citep{NeMe1965} method used in \texttt{R}. The inverse of the negative Hessian at the mode can then be used as the covariance matrix for the asymptotic normal distribution of the MLEs.

The $\alpha$ and $\phi$ parameters are asymptotically orthogonal for a homogeneous Poisson-gamma model \citep{Huzu1950}. A time contraction argument can be used to extend the result to the inhomogeneous case. As discussed in Section 4 of the main paper and visible from (4.3), in the special case where $\tau_c\equiv\tau\;\forall c$, $\theta$ is orthogonal to both $\alpha$ and $\phi$. When carrying out maximum likelihood inference, different model selection criteria such as AIC \citep{Akai1973} and BIC \citep{Schw1978} can be used. Alternatively, one could employ frequentist model averaging methods (see \cite{HjCl2003}, for instance).

The score function and the observed and expected information are provided in the Supplementary Material.
The only pair of parameters which are not asymptotically orthogonal when centres have not been open for the same length of time are $\phi$ and $\theta$. 

\subsection*{Score and observed and expected information}
Here we provide the score function and the observed and expected information, for frequentist inference.

The score function is the gradient of the log-likelihood of the model,
\begin{align*}
&\nabla \ell(\alpha,\phi,\theta|\mathbf{n},\boldsymbol\tau) =\\
&= 
\begin{bmatrix}
C\left(1+\log\frac{\alpha}{\phi}-\psi(\alpha)\right)-\sum_{c=1}^C \left(\frac{\alpha+n^{(\cdot)}_{c}}{\alpha+\phi G(\tau_c;\theta)}+\log\left( G(\tau_c;\theta)+\frac{\alpha}{\phi}\right)-\psi\left(\alpha+n^{(\cdot)}_{c}\right)\right)\tau\\
-C\alpha/\phi+\sum_{c=1}^C\frac{\alpha\left(\alpha+n^{(\cdot)}_c\right)}{\phi(\alpha+\phi G(\tau_c;\theta))}\tau\\
-\sum_{c=1}^C\left[\partial_\theta G(\tau_c;\theta) \left(\frac{\alpha+n^{(\cdot)}_{c}}{G(\tau_c;\theta)+\frac{\alpha}{\phi}}\right)-\sum_{t = 1}^{\tau_c} n_c^{(t)}\left(\frac{\partial_\theta G(t;\theta)-\partial_\theta G(t-1;\theta)}{G(t;\theta)-G(t-1;\theta)}\right)\right]
\end{bmatrix}.
\end{align*}

The observed information matrix is made up of the negative Hessian elements
\begin{align*}
-\partial^2_{\alpha\alpha}\ell(\alpha,\phi,\theta|\mathbf{n},\boldsymbol\tau) = &C\left(\psi'(\alpha)-\frac{1}{\alpha}\right)+\sum_{n=1}^{C}\left\{\frac{\phi G(\tau_c;\theta)- n^{(\cdot)}_c+1}{\alpha+\phi G(\tau_c;\theta)}-\psi'\left(\alpha+n^{(\cdot)}_c\right)\right\},\\
-\partial^2_{\phi\phi}\ell(\alpha,\phi,\theta|\mathbf{n},\boldsymbol\tau) = &-C\alpha/\phi^2+\sum_{c=1}^{C}\frac{\alpha\left(\alpha+2\phi G(\tau_c;\theta)\right)\left(\alpha+n^{(\cdot)}_c\right)}{\phi^2(\alpha+\phi G(\tau_c;\theta))^2},\\
-\partial^2_{\theta\theta}\ell(\alpha,\phi,\theta|\mathbf{n},\boldsymbol\tau) = &\sum_{c=1}^{C}\left[\frac{\left(\alpha+n^{(\cdot)}_c\right)\left\{\partial^2_{\theta\theta}G(\tau_c;\theta)(G(\tau_c;\theta)+\alpha/\phi)-(\partial_\theta G(\tau_c;\theta))^2\right\}}{(G(\tau_c;\theta)+\alpha/\phi)^2}\right.\\
&\left.-\sum_{t=1}^{\tau_c}n_c^{(t)}\frac{H_t \partial^2_{\theta\theta} H_t - (\partial_\theta H_t)^2}{(H_t)^2}\right],\\
-\partial^2_{\alpha\phi}\ell(\alpha,\phi,\theta|\mathbf{n},\boldsymbol\tau) = &\frac{1}{\phi}\left\{C-\sum_{c=1}^C\frac{\alpha^2+2\alpha\phi G(\tau_c;\theta)+\phi G(\tau_c;\theta) n^{(\cdot)}_c}{\left(\alpha+\phi G(\tau_c;\theta)\right)^2}\right\},\\
-\partial^2_{\alpha\theta}\ell(\alpha,\phi,\theta|\mathbf{n},\boldsymbol\tau) = &\sum_{c=1}^{C}\partial_\theta G(\tau_c;\theta)\frac{G(\tau_c;\theta)-n^{(\cdot)}_c/\phi}{\{G(\tau_c;\theta)-\alpha/\phi\}^2},\\
-\partial^2_{\phi\theta}\ell(\alpha,\phi,\theta|\mathbf{n},\boldsymbol\tau) = & -\alpha\sum_{c=1}^{C}\partial_\theta G(\tau_c;\theta)\frac{\alpha+n^{(\cdot)}_c}{\{\alpha+\phi G(\tau_c;\theta)\}^2},
\end{align*}
where $\psi(x) = \Gamma'(x)/\Gamma(x)$ and $H_t = G(t;\theta)-G(t-1;\theta)$ to simplify the notation. Noting that $E{\left[N^{(\cdot)}_c\right]} = \phi G(\tau_c;\theta)$, we obtain the entries of the Fisher information matrix, 
\begin{align*}
E[-\partial^2_{\alpha\alpha}\ell(\alpha,\phi,\theta|\mathbf{N},\boldsymbol\tau)] = &C\left(\psi'(\alpha)-\frac{1}{\alpha}\right)+\sum_{n=1}^{C}\left[\frac{1}{\alpha+\phi G(\tau_c;\theta)}-E{\left\{\psi'\left(\alpha+n^{(\cdot)}_c\right)\right\}}\right],\\
E[-\partial^2_{\phi\phi}\ell(\alpha,\phi,\theta|\mathbf{N},\boldsymbol\tau)] = &\frac{\alpha}{\phi}\sum_{c=1}^{C}\frac{G(\tau_c;\theta)}{\alpha+\phi G(\tau_c;\theta)},\\
E[-\partial^2_{\theta\theta}\ell(\alpha,\phi,\theta|\mathbf{N},\boldsymbol\tau)] = &\sum_{c=1}^{C}\left[\frac{\phi\left\{\partial^2_{\theta\theta}G(\tau_c;\theta)(G(\tau_c;\theta)+\alpha/\phi)-(\partial_\theta G(\tau_c;\theta))^2\right\}}{\phi G(\tau_c;\theta)+\alpha}\right.\\
&\left.-\sum_{t=1}^{\tau_c}n_c^{(t)}\partial^2_{\theta\theta}H_t - \frac{\left(\partial_\theta H_t\right)^2}{H_t}\right],\\
E[-\partial^2_{\alpha\phi}\ell(\alpha,\phi,\theta|\mathbf{N},\boldsymbol\tau)] = &0,\\
E[-\partial^2_{\alpha\theta}\ell(\alpha,\phi,\theta|\mathbf{N},\boldsymbol\tau)] = &0,\\
E[-\partial^2_{\phi\theta}\ell(\alpha,\phi,\theta|\mathbf{N},\boldsymbol\tau)] = &-\alpha\sum_{c=1}^{C}\frac{\partial_\theta G(\tau_c;\theta)}{\alpha+\phi G(\tau_c;\theta)}.
\end{align*}
\section{Bayesian inference}
\label{sec:BayesInf}
For a general model with data $y$, parameter vector $\psi\in\Omega$ and likelihood $f(y|\psi)$, we assign a prior density or mass function to $\psi$, $\pi_0(\psi)$. Inference is based on the posterior distribution, obtained by the Bayes's rule, 
\begin{equation*}
\pi(\psi| y) = \frac{f(y|\psi)\pi_0(\psi)}{\int_{\Omega}f(y|\psi)\pi_0(\psi)\;\mathrm{d}\psi},\quad\psi\in\Omega.
\end{equation*}
Often times, the marginal likelihood of the data $p(y)= \int_{\Omega}f(y|\psi)\pi_0(\psi)\;\mathrm{d}\psi$ is not tractable and so Monte Carlo sampling methods need to be employed to obtain samples from the posterior. Strictly, the marginal likelihood, $p(y)$ is $p(y|M)$ the probability of the data given the choice of model, encapsulated in $f$. Consider, now, a range of models $M_1,\ldots,M_K$ with associated prior probabilities $\pi_0(M_k)$, $k=1,\ldots,K$. Using Bayes's rule, we obtain the posterior model probabilities, up to a proportionality constant,
\begin{align*}
\pi(M_k|y) \propto p(y|M_k)\pi_0(M_k),\quad k=1,\ldots,K.
\end{align*}
\subsection*{Importance sampling}\label{sec:samp}
Multiplying the priors and the likelihood  we obtain the posterior distribution for the parameters up to a proportionality constant. Since the dimension of the parameter space is not large, we can sample from the posterior by the means of importance sampling.

For any function of interest $h(\psi)$,
\begin{align*}
\mathbb{E}[h(\psi)] &= \int_\Omega h(\psi)\pi_p(\psi|y)\;\mathrm{d} \psi= \frac{\int_\Omega h(\psi)\omega(\psi)q(\psi)\;\mathrm{d} \psi}{\int_\Omega f(y|\psi)\pi_0(\psi)\;\mathrm{d} \psi}\\
& \approx \frac{\sum_{b=1}^{B} h\left(\psi^{(b)}\right)\omega\left(\psi^{(b)}\right)}{\sum_{b=1}^{B} \omega\left(\psi^{(b)}\right)},
\end{align*}
where $\psi^{(b)}$, $b=1,\ldots,B$ are samples from a proposal distribution $q$ with unnormalised weights
\begin{equation*}
\omega(\psi) = \frac{f(y|\psi)\pi_0(\psi)}{q(\psi)}.
\end{equation*}

The marginal likelihood may be approximated by
\begin{equation*}
\hat{p}(y) = \frac{1}{B}\sum_{b=1}^{B}\omega\left(\psi^{(b)}\right).
\end{equation*}
This is an unbiased estimate which can be used for model selection or model averaging.

The efficiency of the sampling procedure depends on the choice of proposal distribution $q$ and the may be measured using the effective sample size (ESS),
\begin{equation*}
\mbox{ESS} =\frac{\left(\sum_{b=1}^{B}\omega\left(\psi^{(b)}\right)\right)^2}{\sum_{b=1}^{B}\omega\left(\psi^{(b)}\right)^2}. 
\end{equation*}

If the proposal distribution closely resembles the true posterior, then all the weights will be roughly the same resulting in the ESS being close to $M$. On the other extreme, if the proposal badly captures the posterior and one sample's weight dominates the others, then ESS will be close to one.

If $\psi^{(b)}$ are resampled with replacement with probabilities proportional to the weights, then the resulting sample, say $\{\psi^{(b)}_*\}_{b=1}^{B}$, will have the distribution  approximating $\pi$. The new sample is used when sampling from the predictive distribution to marginalise over the parameter posterior.

\section{Curve-shape prior}\label{app:prior}
The flexible form (4.4) in Section 4 of the main paper, leads to the following prior density for $\tilde{\theta}$,
\begin{equation*}
\pi_0\left(\tilde{\theta}|\kappa,a,b\right) = 
\begin{cases}
t_0\exp\left\{\tilde{\theta}- t_0\exp\{\tilde{\theta}\}\right\}f_{\mathcal{B}}\left(\exp\left\{-t_0\exp\{\tilde{\theta}\}\right\};a,b\right), \quad &\kappa = \infty\\
t_0\exp\{\tilde{\theta}\}\left(1+t_0\exp\{\tilde{\theta}\} /\kappa\right)^{-\kappa-1}f_{\mathcal{B}}\left(\left(1+t_0\exp\{\tilde{\theta}\} /\kappa\right)^{-\kappa};a,b\right), \quad &\kappa\in(0,\infty)
\end{cases},
\end{equation*}
where  $f_{\mathcal{B}}\left(\cdot;a,b\right)$ is a density of a beta variate with shape parameters $a$ and $b$.

\section{Sampling time to completion via model averaging}\label{app:samp_T}

In \cite{LaTa2018}, the time to recruit the required number of patients is sampled by repeatedly simulating the whole system until the condition is satisfied, which is inefficient because each iteration involves a (random) large number of expensive simulations. Additionally, it only provides an approximate distribution due to the discretisation in the time domain; the discretisation of recruitment to monthly increments might also affect the precision of any predictions.  To sample the time to completion exactly, we use the integrated intensity function of the whole trial $\Lambda(t)$. If $T$ is the time to the $m$th arrival of an inhomogeneous Poisson process with integrated intensity $\Lambda(t)$ then \citep{Devr1986} 
\begin{equation*}
\Lambda(T)\sim\mbox{Gamma}(m,1).
\end{equation*}
Given $(\alpha, \beta,\theta)$, we can sample rates the $\lambda^o_c$ for all the centres and construct one realisation of the integrated intensity $\Lambda$ for the whole trial. Then, to obtain a single realisation of $T$, we sample a $\mbox{Gamma}(m,1)$ variate and use an inverse-transform of $\Lambda$ on it. Unless all the centres had been open for the same length of time, the inversion procedure will involve some root-finding algorithm, such as Nelder-Mead \citep{NeMe1965}. As $\Lambda(t)$ in our framework is a monotonically increasing function, the non-linear equation will have a unique solution. Parameter uncertainty can be incorporated into this predictive by using a different sample from the posterior at each iteration. 

Given $C^+$ centres with the first $C$ already opened before the census time and the remaining $C^+-C$ to be open, as well as known centre opening times $t_0^{(c)},\;c=1,\ldots,C^+$, we construct the integrated intensity for modelling the recruitment since the census time $\tau$,
\begin{equation*}
\Lambda(t) = \sum_{c=1}^{C} \lambda^o_c\left\{G\left(t-t_0^{(c)};\theta\right) - G(\tau_c;\theta)\right\}
			+\sum_{c=C+1}^{C^+} \lambda^o_cG\left(t-t_0^{(c)};\theta\right)\chi_{\left\{t>t_0^{(c)}\right\}},\quad t\geq\tau,
\end{equation*}
where $\chi_{\{\cdot\}}$ is the indicator function and
\begin{equation}
\lambda^o_c|\alpha,\phi,\theta,\mathbf{n}\sim\begin{cases}
\mbox{Gamma}\left(\alpha+n^{(\cdot)}_{c}, \alpha/\phi+G(\tau_c;\theta)\right),\quad& c = 1,\ldots,C\\
\mbox{Gamma}\left(\alpha, \alpha/\phi\right),\quad& c = C+1,\ldots,C^+
\end{cases}.\label{eqn:lam0}
\end{equation}
The algorithm below outlines the sampling procedure to obtain the distribution of the time needed to recruit the target number of patients $m$.

\IncMargin{1em}
\begin{algorithm}
\SetKwData{Left}{left}\SetKwData{This}{this}\SetKwData{Up}{up}
\SetKwFunction{Union}{Union}\SetKwFunction{FindCompress}{FindCompress}
\SetKwInOut{Input}{input}\SetKwInOut{Output}{output}

\Input{Models $\Mb_1,\ldots,\Mb_k$ with posterior probabilities $\pi(\Mb_1|\bm{n}),\ldots,\pi(\Mb_K|\bm{n})$ and posterior samples from each model, number of samples from the predictive $B$, target number of recruitments $m$ }
\Output{Distribution of the time to completion $\left\{T^{(b)}\right\}_{b=1}^B$}
\BlankLine

\For{$b\leftarrow 1$ \KwTo $B$}{
Sample $\Mb^{(b)}\sim\pi(\Mb_k|\bm{n}) $\;
Sample $(\alpha,\phi,\theta)^{(b)}\sim\pi(\alpha,\phi,\theta|\Mb^{(b)},\bm{n})$\;
Sample rates $\lambda^o_c|(\alpha,\phi,\theta)^{(b)}$ from distributions (\ref{eqn:lam0}) and construct $\Lambda^{(b)}(t)$\;
Sample $\tilde{T}\sim\mbox{Gamma}(m,1)$ and solve $\Lambda^{(b)}(T) = \tilde{T}$\;
Set $T^{(b)} = T$\;
}
\caption{Model-averaged time to completion sampling}\label{alg:time}
\end{algorithm}\DecMargin{1em}
\newpage
\section{Additional details from the simulation study and data analysis}
\subsection{Simulation study}
\begin{figure}
\centering
\includegraphics[width=0.8\textwidth]{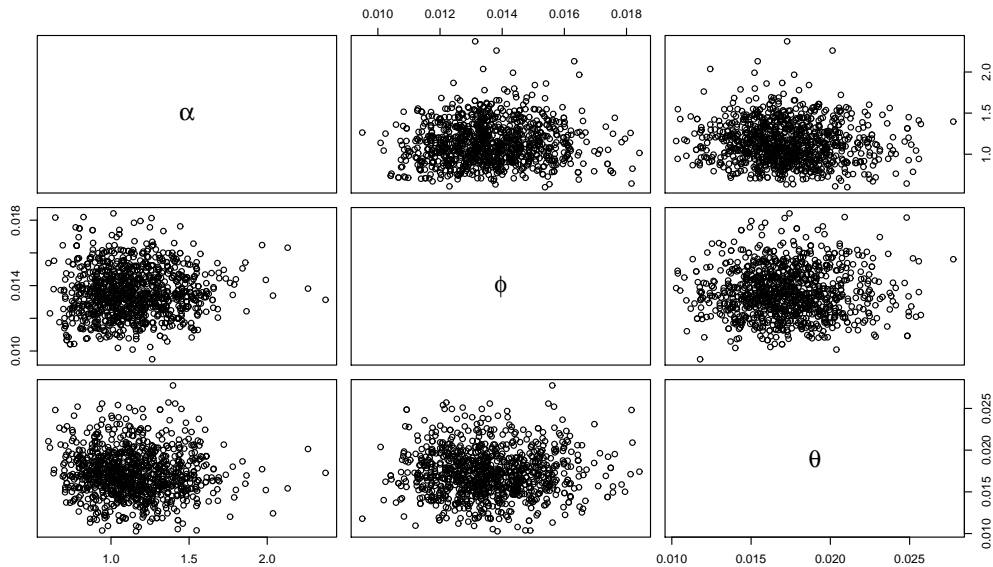}
\caption{Matrix scatterplot of the parameter posterior of the model with highest posterior probability.}
\label{fig:sim_post}
\end{figure}

Figure \ref{fig:sim_post}  shows the plots of posterior samples of the model. The three parameters are close to orthogonal as discussed in Sections 4 and 5 of the paper, and this approximate independence was also observed in the posteriors of other models.

\begin{figure}
    \centering
    \begin{minipage}{0.45\textwidth}
        \centering
        \includegraphics[width = 0.8\textwidth]{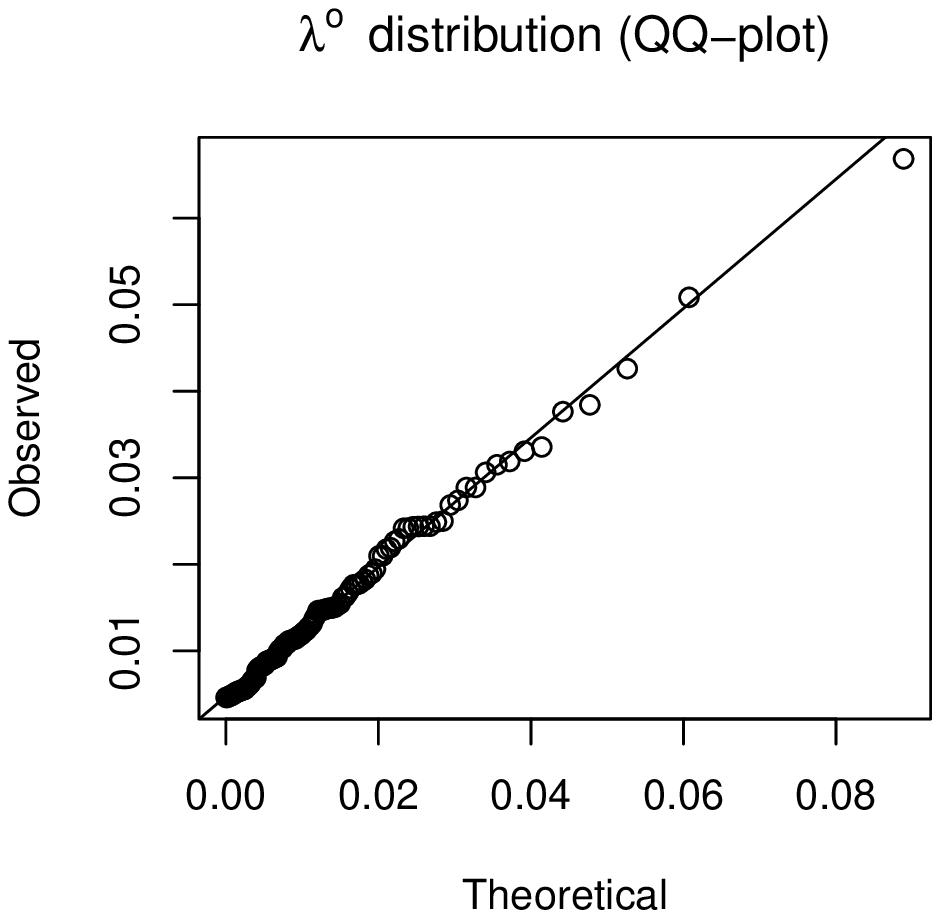}
\caption{Re-estimated $\lambda^o_c$ expectations compared to Gamma$\left(\hat\alpha,\hat{\alpha}/\hat{\phi}\right)$ distribution.}
\label{fig:sim_diag1}
    \end{minipage}\hfill
    \begin{minipage}{0.45\textwidth}
        \centering
        \includegraphics[width = 0.8\textwidth]{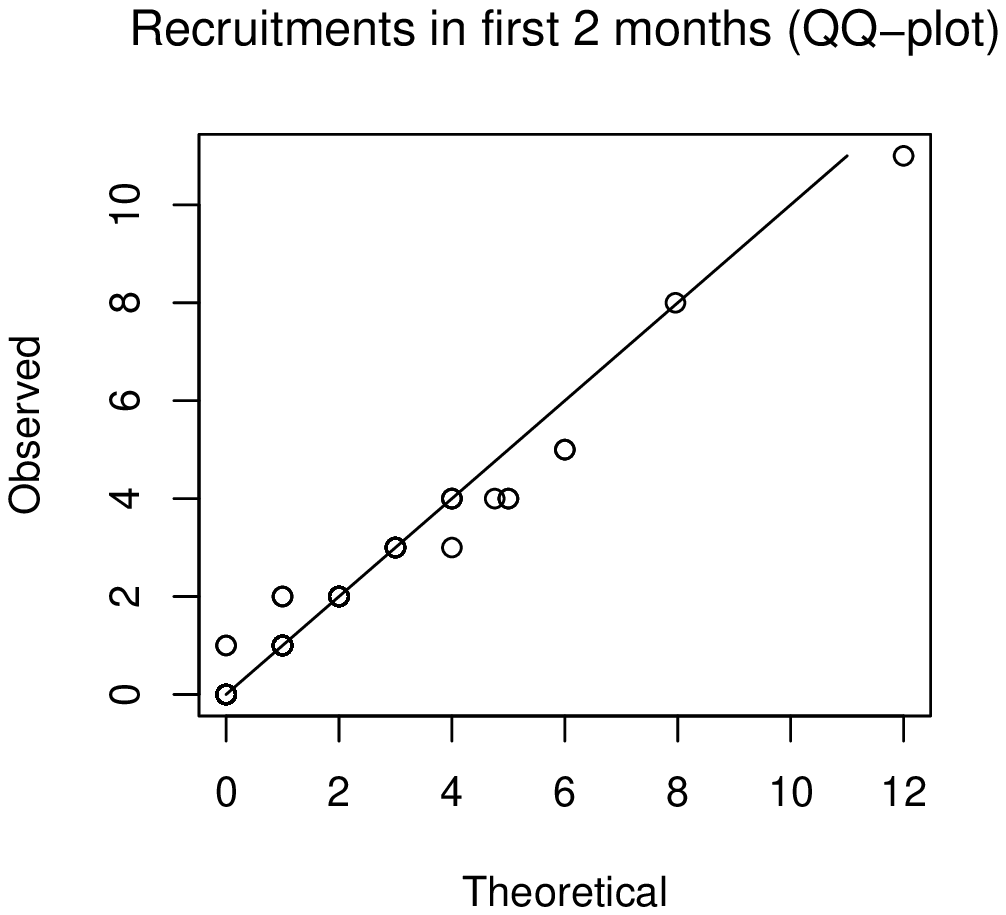}
\caption{Observed recruitments compared to the theoretical negative binomial distribution.}
\label{fig:sim_diag2}
    \end{minipage}
\end{figure}

\begin{figure}
    \centering
    \begin{minipage}{0.45\textwidth}
        \centering
        \includegraphics[width = 0.8\textwidth]{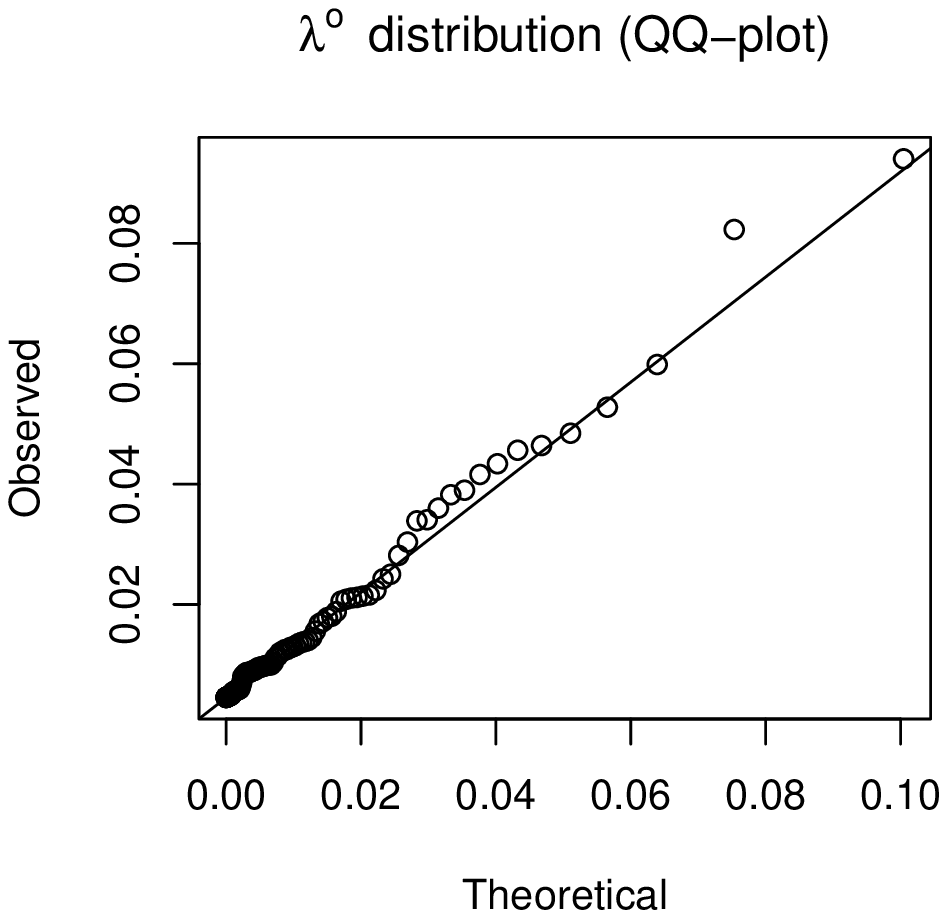}
\caption{Re-estimated $\lambda^o_c$ expectations compared to Gamma$\left(\hat\alpha,\hat{\alpha}/\hat{\phi}\right)$ distribution; true random-effect distribution is a mixture with $\mathbb{E}[\lambda^o_c] = 0.01$.}
\label{fig:sim_bim1_diag1}
    \end{minipage}\hfill
    \begin{minipage}{0.45\textwidth}
        \centering
        \includegraphics[width = 0.8\textwidth]{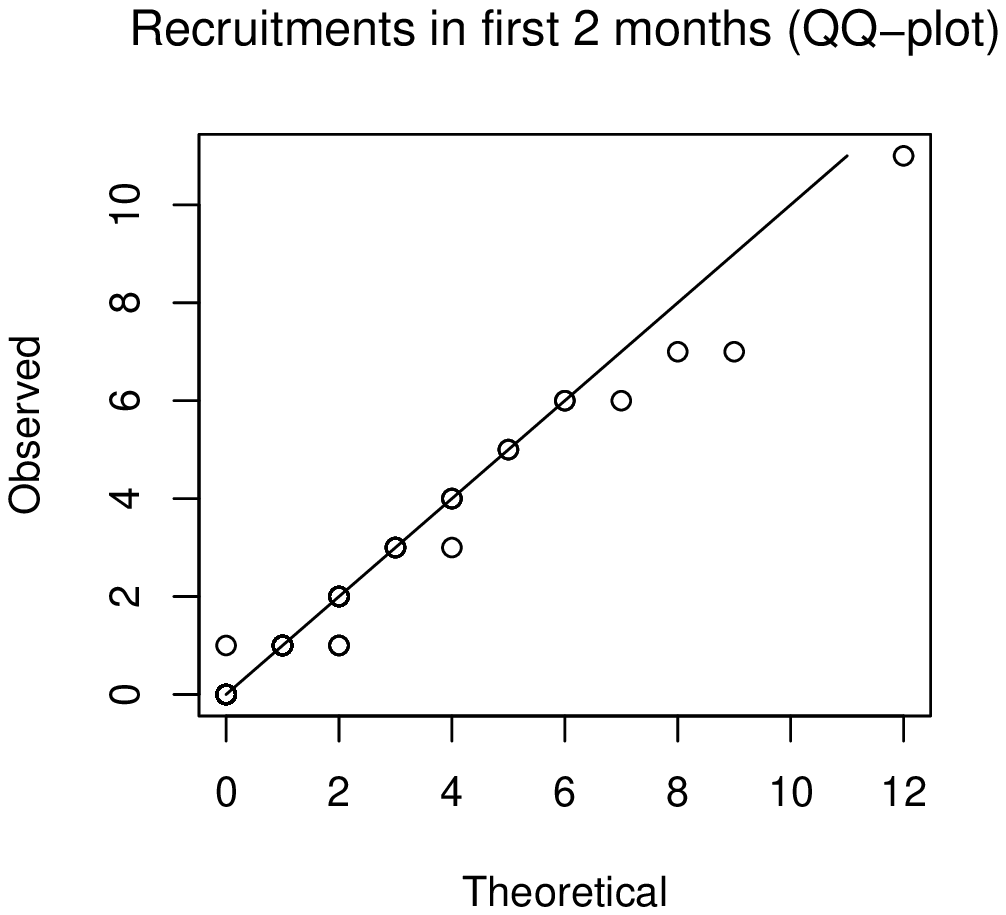}
\caption{Observed recruitments compared to the theoretical negative binomial distribution; true random-effect distribution is a mixture with $\mathbb{E}[\lambda^o_c] = 0.01$}
\label{fig:sim_bim1_diag2}
    \end{minipage}
\end{figure}

\begin{figure}
    \centering
    \begin{minipage}{0.45\textwidth}
        \centering
        \includegraphics[width = 0.8\textwidth]{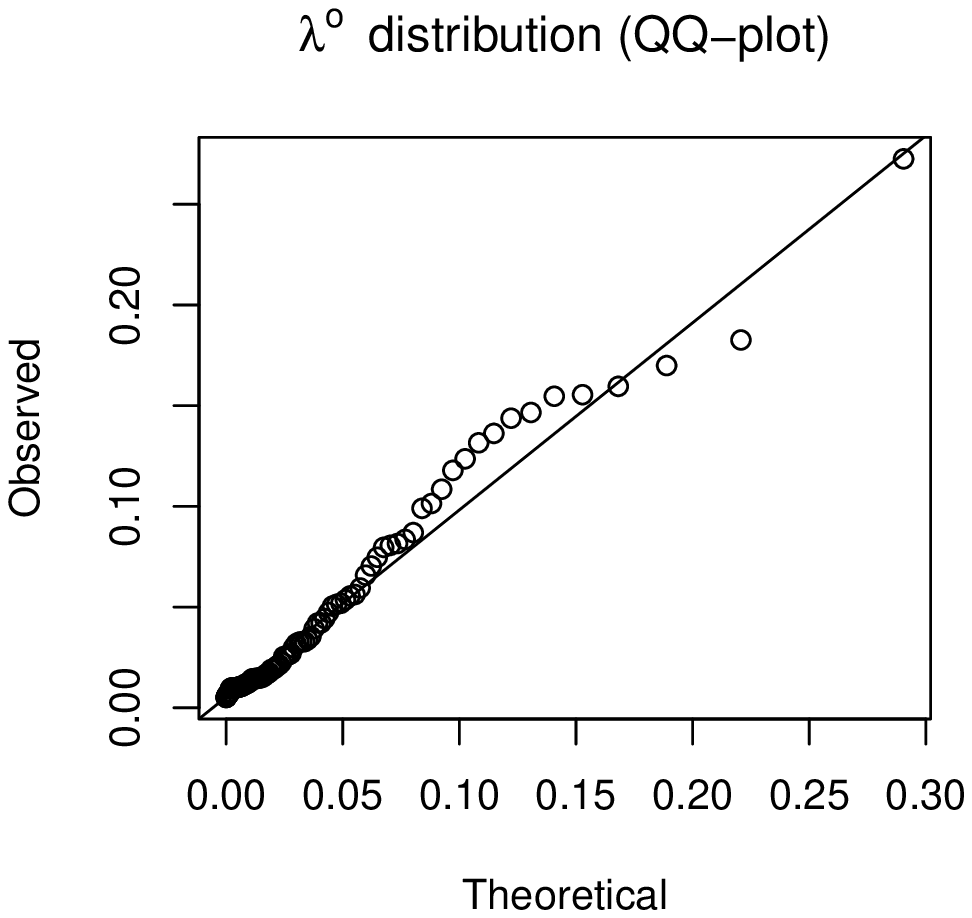}
\caption{Re-estimated $\lambda^o_c$ expectations compared to Gamma$\left(\hat\alpha,\hat{\alpha}/\hat{\phi}\right)$ distribution; true random-effect distribution is a mixture with $\mathbb{E}[\lambda^o_c] = 0.03$.}
\label{fig:sim_bim3_diag1}
    \end{minipage}\hfill
    \begin{minipage}{0.45\textwidth}
        \centering
        \includegraphics[width = 0.8\textwidth]{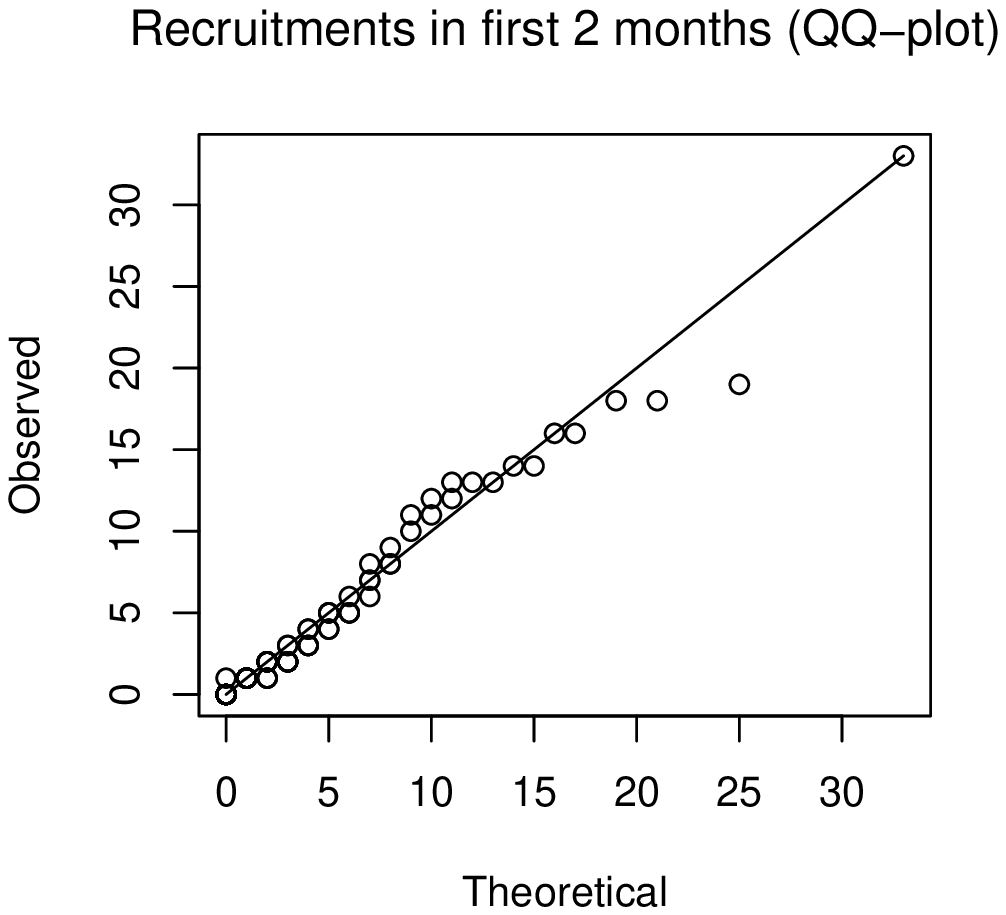}
\caption{Observed recruitments compared to the theoretical negative binomial distribution; true random-effect distribution is a mixture with $\mathbb{E}[\lambda^o_c] = 0.03$}
\label{fig:sim_bim3_diag2}
    \end{minipage}
\end{figure}

Figure \ref{fig:sim_diag1} shows a QQ-plot of the hierarchical gamma distribution compared to the posterior means of the random effects. The approximately straight line indicates that generating rates for newly opened centres from the gamma distribution will be consistent with what has been observed thus far.  Figure \ref{fig:sim_diag2}  shows a QQ-plot of the theoretical, negative binomial distribution of recruitments in the first 2 months compared the observed distribution ($t_*=60$). The theoretical distribution used the posterior means of the parameters, and the prior random effect distribution was used. The straight line shows that the model can predict the recruitment in the first two months of a centre sufficiently well. In practice, the two diagnostics would indicate that the mixing gamma distribution is sufficient and that the model is capable of accurately predicting recruitments in the early days of a new centre.
Figures \ref{fig:sim_bim1_diag1}, \ref{fig:sim_bim1_diag2}, \ref{fig:sim_bim3_diag1} and \ref{fig:sim_bim3_diag2} show the diagnostic plots for models fit to simulated datasets at the census $t=360$ with the true random-effect distribution being a mixture. For $E[\lambda^o_c] = 0.01$, the relationship is close to linear and is reflected in the reasonably accurate predictions shown in the article. The QQ-plots for $E[\lambda^o_c] = 0.03$ show stronger non-linearity and informing us of the potential misspecification, thus showing that the diagnostics can be used to validate the model.

\begin{figure}
\begin{subfigure}{.5\textwidth}
  \centering
  \includegraphics[width=.9\textwidth]{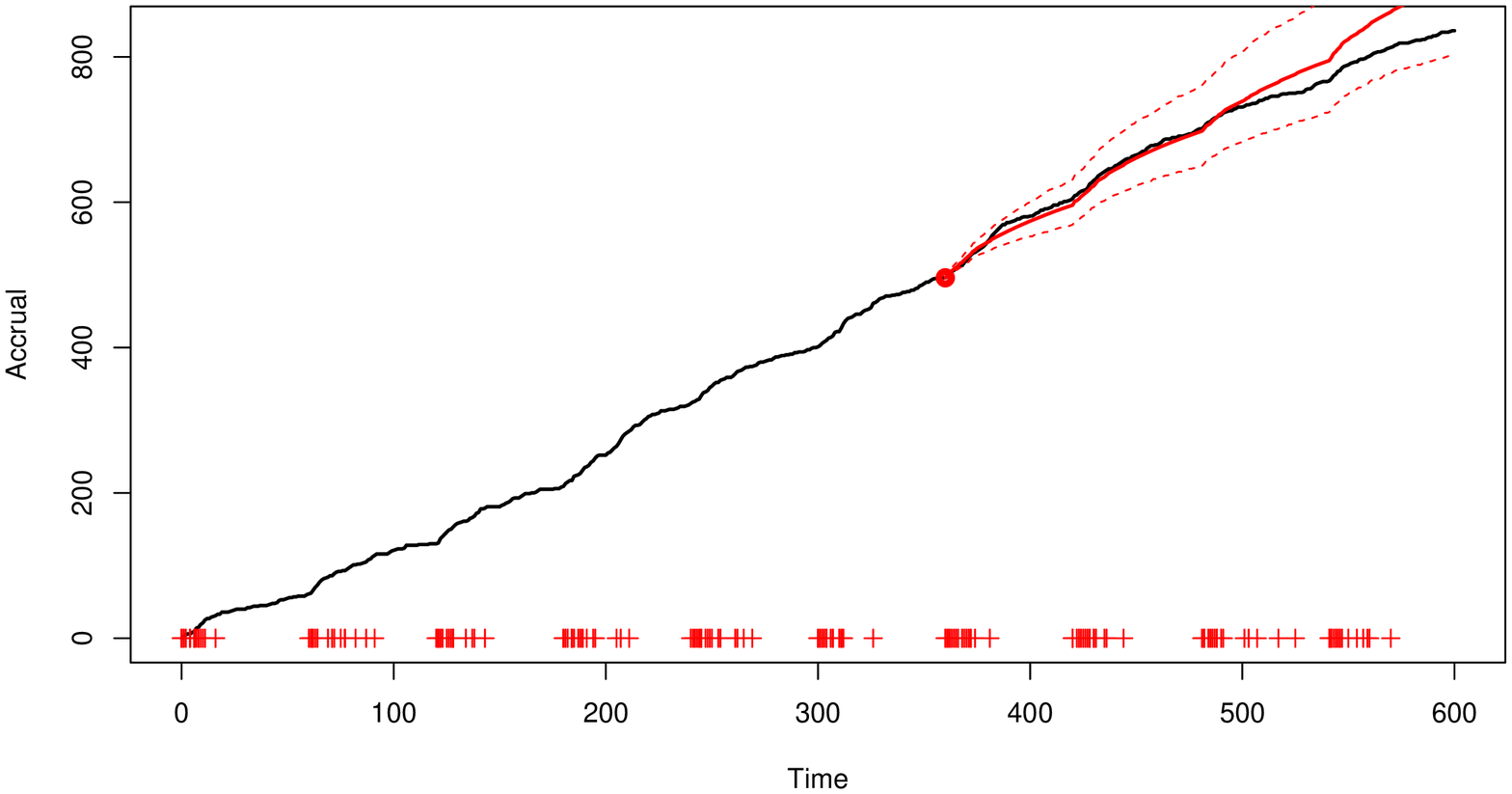}
  \caption{\normalsize{ Clumped openings, $\mathbb{E}[\lambda^o_c] = 0.01$;\\
  $p$-value = 0.666}}
  \label{fig:sim_bim2}
\end{subfigure}
\begin{subfigure}{.5\textwidth}
  \centering
  \includegraphics[width=.9\textwidth]{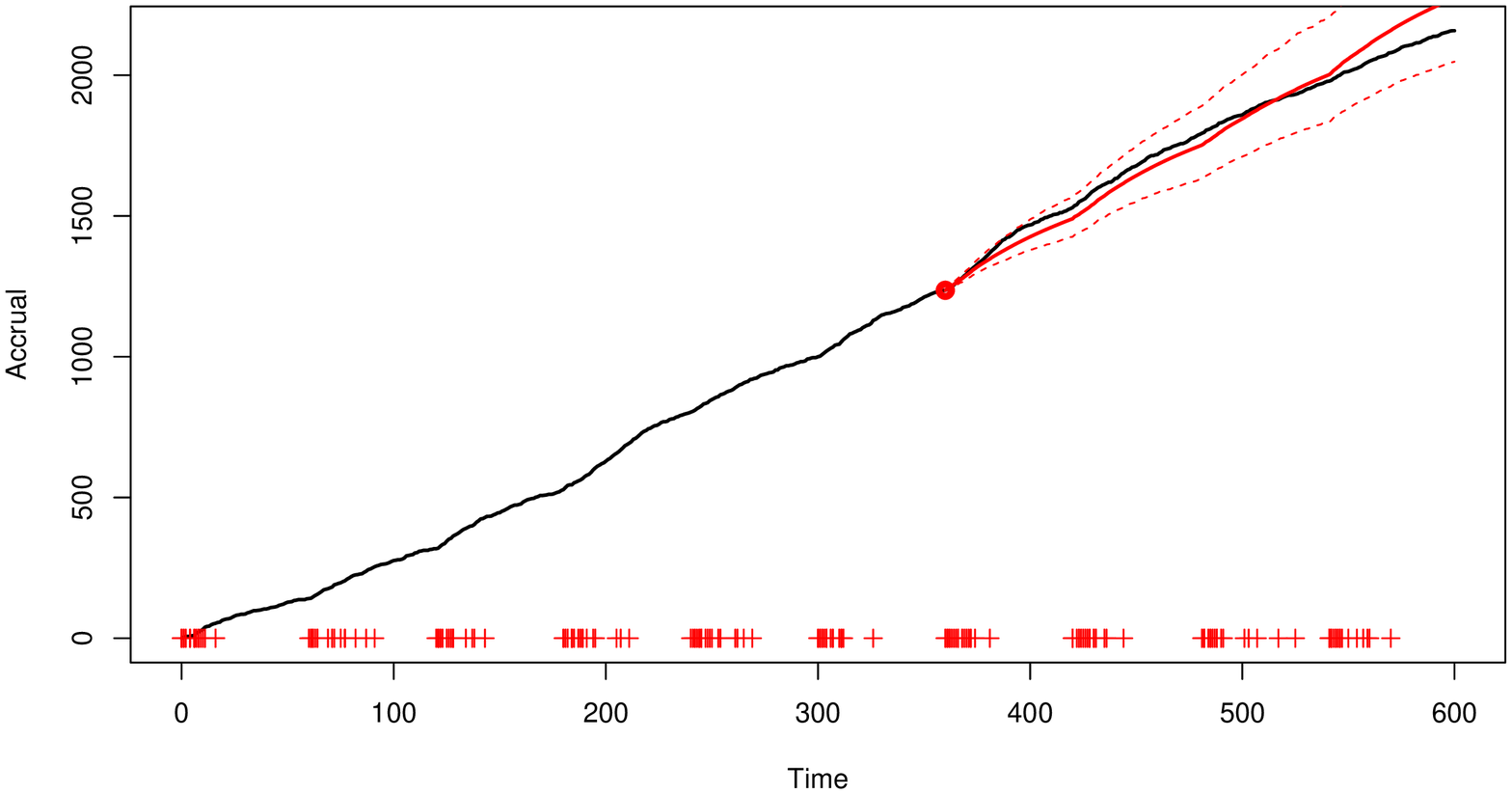}
  \caption{\normalsize{ Clumped openings, $\mathbb{E}[\lambda^o_c] = 0.03$;\\
  $p$-value = 0.312}}
  \label{fig:sim_bim4}
\end{subfigure}
\caption{Accruals (black, solid) with predictive means (red, solid) and 95\% prediciton bands (red, dashed) when the true random-effect distribution is a mixture, in various scenarios.  Prediction bands are based on the $2.5\%$ and $97.5\%$ quantiles. The ``+'' symbols on the abscissa indicate centre opening times.}
\label{fig:sim_bim_clump}
\end{figure}

Figures \ref{fig:sim_bim2} and \ref{fig:sim_bim4} show examples of recruitment predictions when the random effects have a mixture distribution and the centre opening times are ``clumped'' together. The clumping accentuates the effect of the misspecification; the fitted model relies on the ``incorrect'' prior gamma distribution when simulating rates for unopened centres.

\begin{figure}
\begin{subfigure}{.5\textwidth}
  \centering
  \includegraphics[width=.9\textwidth]{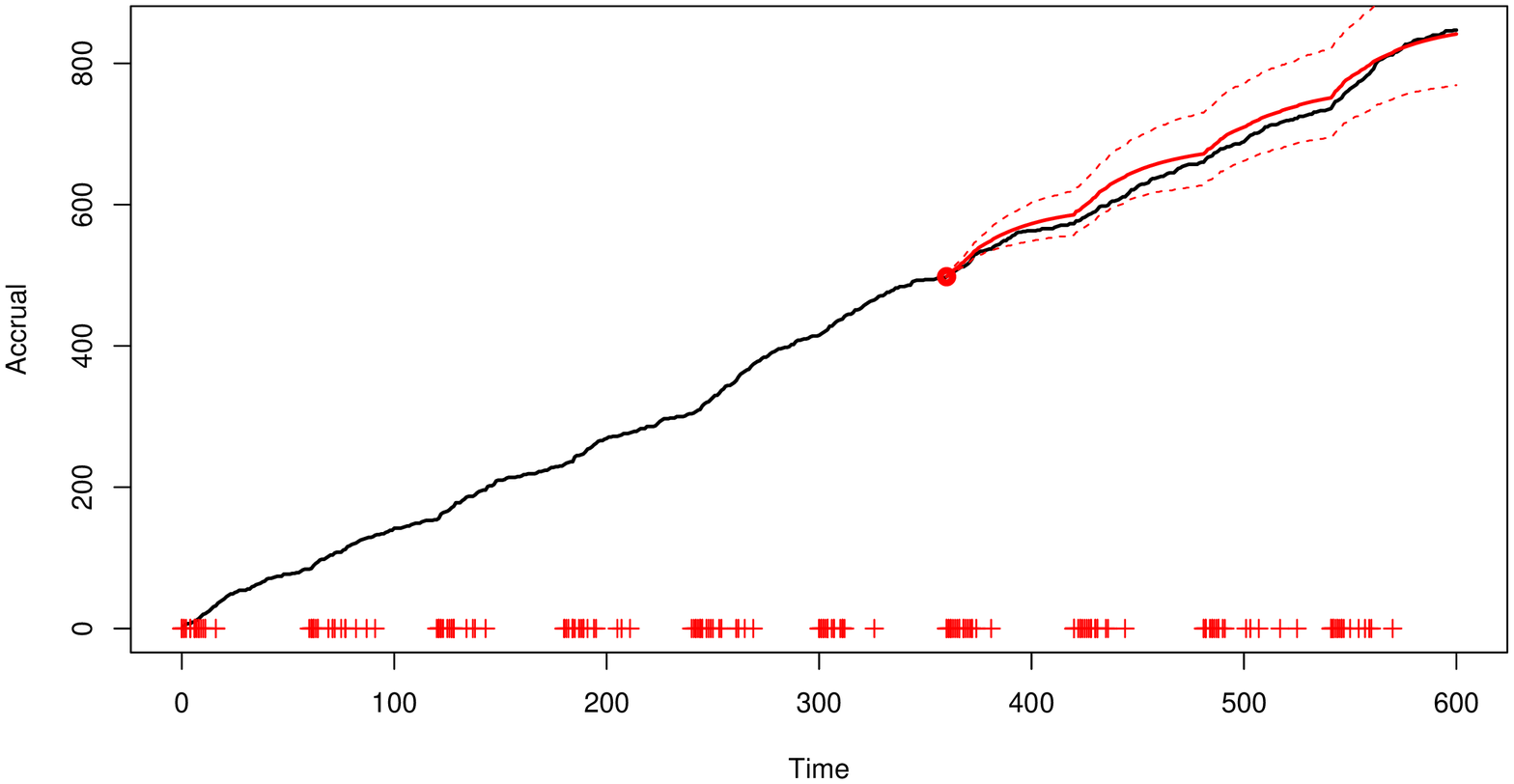}
  \caption{\normalsize{ Clumped openings, $\mathbb{E}[\lambda^o_c] = 0.01$;\\
  $p$-value = 0.294}}
  \label{fig:sim_weib2}
\end{subfigure}
\begin{subfigure}{.5\textwidth}
  \centering
  \includegraphics[width=.9\textwidth]{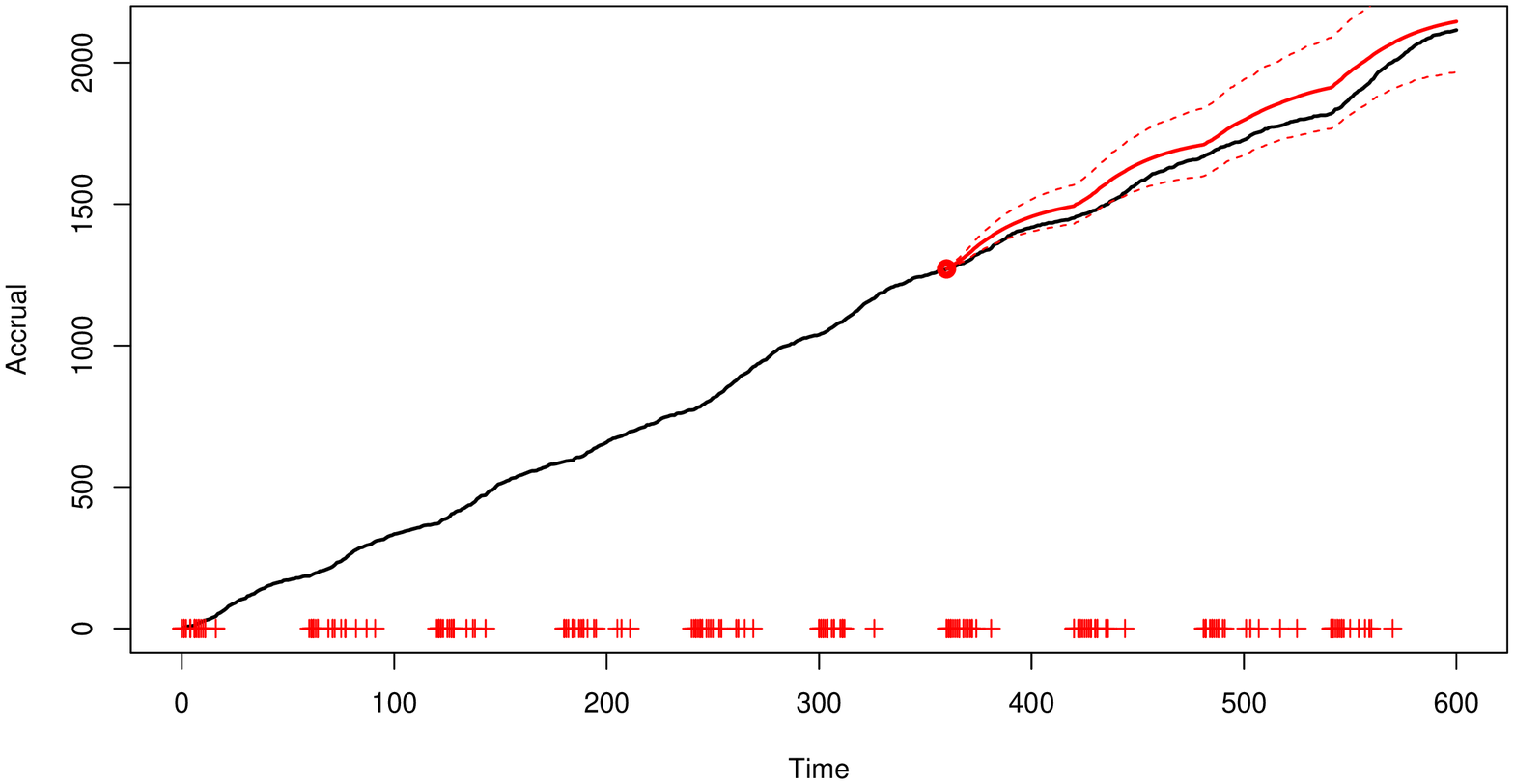}
  \caption{\normalsize{ Clumped openings, $\mathbb{E}[\lambda^o_c] = 0.03$;\\
  $p$-value = 0.064}}
  \label{fig:sim_weib4}
\end{subfigure}
\caption{Accruals (black, solid) with predictive means (red, solid) and 95\% prediciton bands (red, dashed) when the true intensity shape is Weibull and opening times are clumped, with two different values for $\mathbb{E}[\lambda_c^o]$.  Prediction bands are based on the $2.5\%$ and $97.5\%$ quantiles. The ``+'' symbols on the abscissa indicate centre opening times.}
\label{fig:sim_weib_clump}
\end{figure}
 
Figures \ref{fig:sim_weib2} and \ref{fig:sim_weib4} show the predictions made when using data simulated from a Weibull-shape intensity with centre opening times clumped together. 
With repeated simulations, we found a consistent correspondence between linear QQ-plots and accurate predictions.

\subsection{Data analysis}

\begin{figure}
	    \centering
	    \includegraphics[width = 0.6\textwidth]{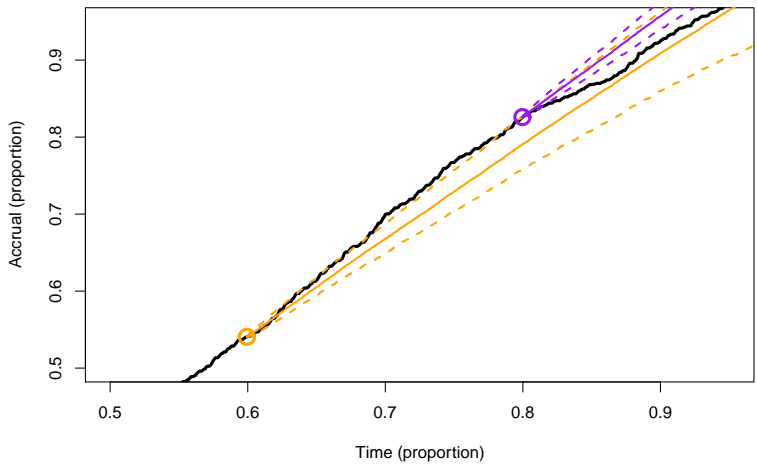}
	    \caption{Accrual predictions, zoomed-in to focus on the unexpected surge in recruitment at around the time of $0.7$. Only interim forecasts from times $0.6$ and $0.8$ are shown.}
	    \label{fig:zoom}
	\end{figure}

In the dataset examined in Section 7 of the main paper, we encountered an unexpected surge in recruitments at a global scale. Figure \ref{fig:zoom} shows the accrual along with 2 sets of forecasts, focusing on the surge at a time of around $0.7$. Once this has been observed, and forward predictions are needed, one possibility is modelling this as a global surge in recruitment; that is, during the period between $0.6$ and $0.75$ all recruitment rates are multiplied by $\exp\{\beta\}$ for some unknown $\beta$, which would be an extra parameter to be estimated via importance sampling. 

\section{Stochastic centre-initiation times }

\begin{figure}
        \centering
        \includegraphics[width=.95\textwidth]{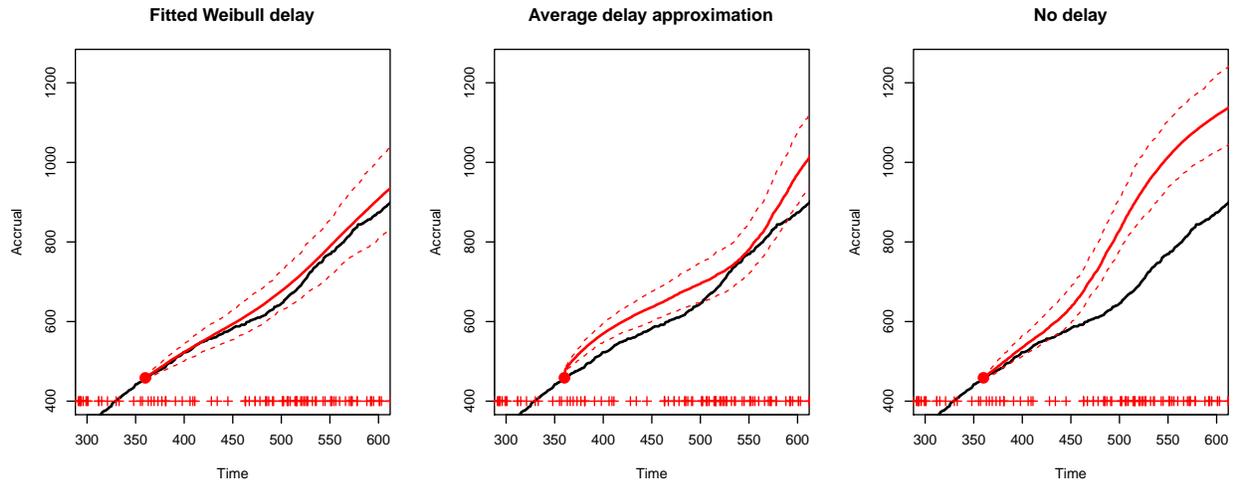}
  \caption{Comparison of predictions for recruitment data with stochastically-delayed centre-initiation times. Three modelling approaches are considered: correct Weibull-distributed delay fitted (left); constant, historical average delay added to each initiation time (centre); and no delay considered in predictions (right). }
  \label{fig:delay_compare}
    
\end{figure}

The framework, as presented in the main paper, is conditioned on the set of initiation times both for clarity of presentation and because it is the methodological contribution from the paper. In practice, the exact future initiation times would be unknown; instead, the practitioners would have proposed initiation schedules, contingency plans and recruitment data up to the census time. Here we present a simulation study similar to that in Section 6 of the main paper which illustrates how a stochastic centre-initiation model can be seamlessly incorporated. The centres are not initiated exactly on schedule but, instead, there is a Weibull-distributed initiation delay for each centre. Following information provided to us from a large meta-analysis, we set the Weibull parameters such that the $5$th and $95$th percentiles are 10 and 322 days respectively; the median delay is 90 days. At the census, the observed day-censored delays are used for maximum-likelihood estimation of the Weibull parameters; additionally centres which were planned to initiate before the census but did not do so  contribute with a censored likelihood. The estimates are then used in the Monte Carlo simulations. Figure \ref{fig:delay_compare} compare the predictions under three different approaches; (i) the correct Weibull distribution for delays (with parameters estimated from the data), (ii) a constant, avergae delay taken to be the sample mean of the observed delays, and (iii) an assumption of no delays. It is clear that assuming no future delays given historical evidence of the contrary leads to poor forecasts. However, even very simple delay predictions based on the empirical average can achieve desirable forecasts. Of course, fitting the true model results in predictions which capture the truth extremely well. This illustrates that our site-level prediction method can be easily combined with site-initiation models.
\unappendix

\end{document}